\documentclass[prd,10pt,nofootinbib,superscriptaddress]{revtex4-2}
\usepackage{bm}
\usepackage{graphicx}
\usepackage{hyperref}
\usepackage{comment}
\usepackage{color}
\usepackage{orcidlink}
\usepackage{ulem}
\usepackage{subfigure}
\hypersetup{colorlinks=true,linkcolor=red,citecolor=blue}
\usepackage{xcolor}
\usepackage{amsmath,amsfonts,amssymb,amstext}
\usepackage{multirow}
\usepackage{float}
  \usepackage{subcaption}
\usepackage{cases}
\usepackage{epsfig}%
\definecolor{darkblue}{HTML}{000000}
\begin{document}
\title{Observational and Thermodynamic aspects of one-dimensional Dark Energy EoS parametrization models}
 
\author{Anirban Chatterjee \orcidlink{0000-0002-8904-7133}}
\email{Corresponding author: anirbanchatterjee@nbu.edu.cn \& iitkanirbanc@gmail.com}
\author{Yungui Gong \orcidlink{0000-0001-5065-2259}} 
\email{gongyungui@nbu.edu.cn}
\affiliation{Institute of Fundamental Physics and Quantum Technology, Department of Physics, School of Physical Science and Technology, Ningbo University, Ningbo, Zhejiang 315211, China}

\begin{abstract}
We investigate the observational and thermodynamic viability of  Gong--Zhang (GZ) Type~I (GZ1) and Type~II (GZ2) dark--energy parametrizations using late--time cosmological probes including Type~Ia supernovae (Union3, Pantheon+SH0ES, and DES--SN5YR), DESI baryon acoustic oscillations, cosmic chronometer $H(z)$ measurements, and growth--rate data. Using Bayesian Markov Chain Monte Carlo analysis together with Akaike and Bayesian information criteria, we show that both parametrizations provide observationally consistent and phenomenologically competitive late-time alternatives to $\Lambda$CDM, while the  GZ2 model generally provides tighter constraints and reduced parameter degeneracies. The reconstructed evolution of the dark--energy equation of state and coincidence parameter demonstrates that both models recover the standard matter--dominated behaviour at high redshift while producing controlled late--time deviations from the cosmological--constant scenario. A complementary cosmographic, sound--speed, and growth analysis further confirms stable late--time accelerated expansion together with physically viable perturbative behaviour. Finally, using configuration entropy as a thermodynamic probe, we show that the entropy--production rate sensitively captures the influence of dynamical dark energy on late--time structure formation while remaining consistent with standard early--time cosmology.
\end{abstract}

\maketitle

\section{Introduction}
\label{sec:Intro}
 
Observational evidence accumulated over the past two decades has firmly established that the universe is presently undergoing a phase of accelerated expansion \cite{ref:Riess98,ref:Perlmutter}. Within the standard cosmological paradigm, this phenomenon is attributed to a cosmological constant $\Lambda$, leading to the concordance $\Lambda$CDM model. Despite its remarkable empirical success in describing a wide range of cosmological observations, $\Lambda$CDM continues to face persistent theoretical and phenomenological challenges, most notably the cosmological--constant problem \cite{Zlatev:1998tr}, fine-tuning problem \cite{Martin:2012bt}, and the growing tensions among late--time measurements of the Hubble parameter \cite{ht1,ht2,Pourtsidou:2016ico} and large--scale structure \cite{SDSS:2005xqv}. These open issues motivate the exploration of dynamical dark-energy models \cite{Tsujikawa:2012hv} that extend $\Lambda$CDM while maintaining phenomenological consistency. Scalar-field scenarios provide a natural realization of late-time acceleration: quintessence models \cite{Peccei:1987mm,Ford:1987de, Chatterjee:2024duy, Chatterjee:2025htd, Copeland:1997et} employ slowly rolling potentials, whereas $k$ essence models \cite{Fang:2014qga,ArmendarizPicon:1999rj,ArmendarizPicon:2000ah, Bandyopadhyay:2017igc,Bandyopadhyay:2018zlz,Bandyopadhyay:2019vdd,Chatterjee:2022uyw} rely on non-canonical kinetic terms, with both classes generating effective negative pressure. Alternatively, late-time acceleration may arise from modifications of gravity itself, including  f(R), scalar–tensor, Gauss–Bonnet, and brane-world theories \cite{fr1,fr2,fr3,fr4,fr5,fr6,fr7}, which offer rich phenomenology beyond the cosmological-constant paradigm. A complementary alternative is provided by interacting field–fluid scenarios, where energy–momentum exchange between dark energy and pressureless matter modifies background and perturbative dynamics, potentially alleviating the cosmic coincidence problem while remaining observationally viable \cite{Chatterjee:2021ijw,Chatterjee:2021hhj,Hussain:2022osn,Bhattacharya:2022wzu,Hussain:2023kwk,Chatterjee:2023uga}.\\

A minimal and widely adopted strategy to model dynamical dark energy is through phenomenological parametrizations of the equation--of--state parameter $w(z)$ \cite{Gao:2024ily,Bandyopadhyay:2019ukl, Yang:2018uae, Gong:2005de}. Such parametrizations allow for controlled departures from a pure cosmological constant without committing to a specific microscopic realization. Among the various proposals, one--dimensional parametrizations play a special role owing to their simplicity, stability, and direct interpretability in terms of observational constraints. In this context, the Gong--Zhang parametrizations \cite{Gong:2005de} constitute a particularly well--behaved class.
They are explicitly constructed to recover an effective matter–dominated expansion behaviour and to generate smooth, monotonic deviations from $\Lambda$CDM only at late times, thereby preserving early--Universe physics while enabling observationally testable modifications of cosmic acceleration. This makes them especially suitable for isolating late--time dynamics using low--redshift probes alone.\\

While most observational studies of dark energy focus primarily on background expansion or linear growth observables \cite{Sagredo:2018ahx, Marcondes:2016reb, Kazantzidis:2018rnb}, cosmic acceleration also leaves a cumulative imprint on the formation of cosmic structures. From a physical standpoint, this motivates the introduction of diagnostics that capture not only instantaneous expansion or growth rates, but also the integrated history of gravitational clustering. Configuration entropy, inspired by Shannon information entropy \cite{Bekenstein:2008smd,Pandey:2017tgy,Das:2018shd,Shannon:1948} and rooted in thermodynamic considerations, provides such a global measure. It quantifies the degree of spatial inhomogeneity in the matter distribution and monotonically decreases as gravitational instability amplifies initially small density perturbations. As structures form and the Universe becomes increasingly clumpy, configuration entropy dissipates, encoding the irreversible flow of information associated with structure formation.\\

In this work, we perform a comprehensive observational, dynamical, and thermodynamic analysis of the Gong--Zhang \cite{Gong:2005de} Type~I (GZ1) and Type~II (GZ2) dark--energy parametrizations using three independent Type~Ia supernova dataset combinations along with BAO, OHD, and $f\sigma_8$ observations \cite{DESI:2024mwx,Scolnic:2021amr,Rubin:2023jdq,DES:2024jxu,DES:2025sig,Gong:2025hoy,Gao:2025ozb,Lu:2024hvv}. We begin by presenting the marginalized one-- and two--dimensional posterior distributions for the parameter set $\{\Omega_m,\,h,\,h\,r_d,\,\omega_0,\,\sigma_8\}$ together with the corresponding correlation matrices in order to identify the dominant degeneracy directions among the matter sector, the cosmological distance scale, and the dark--energy dynamics. We then report the best--fit cosmological parameters along with the Akaike and Bayesian information criteria and perform a Jeffreys--scale model comparison relative to $\Lambda$CDM. The marginalized $\Omega_m$--$\omega_0$ and $\Omega_m$--$\sigma_8$ constraints are further analyzed to characterize the late--time dynamical and clustering behaviour of the Gong--Zhang parametrizations and their departures from the standard $\Lambda$CDM scenario. Beyond the background expansion history, we investigate the kinematic and dynamical implications of the Gong--Zhang models through a detailed cosmographic analysis together with the reconstructed evolution of the dark--energy equation of state $\omega(z)$ and the coincidence parameter $r_{mc}(z)=\frac{\Omega_m(z)}{\Omega_{\rm DE}(z)}$. These quantities provide a direct characterization of the transition from the present accelerated epoch to the standard matter--dominated regime at high redshift and allow us to examine how the Gong--Zhang parametrizations preserve the standard early--time cosmological behaviour while generating controlled late--time deviations from $\Lambda$CDM. We further present the corresponding sound--speed analysis in order to identify the regions satisfying the physical stability condition $0<c_s^2<1$, thereby ensuring perturbative stability and causal propagation of dark--energy fluctuations.\\

To further probe the physical implications of dynamical dark energy, we examine its impact on the formation of cosmic structures by studying the relative deviation of the linear growth rate with respect to $\Lambda$CDM, thereby quantifying how the Gong--Zhang parametrizations modify the growth of matter perturbations while accurately reproducing the standard early--time growth behaviour and inducing a mild suppression of structure formation at late times. Beyond conventional growth diagnostics, we employ configuration entropy as a physically motivated thermodynamic probe of structure formation, capable of capturing the cumulative and irreversible aspects of gravitational clustering. By analyzing the evolution of the configuration--entropy production rate and its deviation from the $\Lambda$CDM prediction across different SN+BAO+OHD+$f\sigma_8$ dataset combinations, and by connecting this behaviour to the background expansion and growth history of the Gong--Zhang models, we establish a unified framework linking late--time cosmic acceleration, structure growth, and entropy flow. This combined dynamical and thermodynamic approach provides a deeper physical motivation for studying one--dimensional dark--energy parametrizations and offers a robust criterion for assessing the viability of the Gong--Zhang scenario beyond conventional background observables.\\

The paper is organized as follows. In Sec.~\ref{sec:data} we describe the observational datasets, cosmological model framework, and statistical methodology. Sec.~\ref{sec:obs} presents the observational constraints, parameter correlations, and information--criteria analysis. In Sec.~\ref{sec:CE} we investigate the evolution of configuration entropy and its sensitivity to the Gong--Zhang dynamics.
Finally, Sec.~\ref{sec:con} summarizes our main conclusions and outlook.\\

\section{Observational data, cosmological model framework, and statistical methodology}
\label{sec:data}

In this section, we describe the observational datasets, cosmological framework,
and statistical methodology adopted to constrain the Gong--Zhang Type~I (GZ1)
and Type~II (GZ2) dark--energy parametrizations. Throughout this work we assume
a spatially flat FLRW Universe and use natural units with $c=1$.

% ============================================================
\subsection{Observational datasets}
% ============================================================

We constrain the Gong--Zhang parametrizations using a combination of late--time
cosmological probes sensitive to both the background expansion history and the
growth of matter perturbations. The analysis is intentionally performed without
including any direct CMB likelihood in order to maintain a phenomenological late–time framework. Since the Gong--Zhang parametrizations are primarily intended as effective late--time descriptions of dark energy, we restrict the present analysis to low--redshift cosmological probes and defer a full early--Universe and CMB consistency analysis to future work.

\medskip
\noindent\textbf{(a) Baryon acoustic oscillations:}
We use BAO measurements from the first-year DESI observations
\cite{DESI:2024mwx}, spanning the redshift range $0.1<z<4.2$.
The BAO information is incorporated through the compressed observables
$\frac{D_M(z)}{r_d}$ and $\frac{D_H(z)}{r_d}$,
where $D_M(z)$ denotes the transverse comoving distance and
$D_H(z)=1/H(z)$ (in units with $c=1$) represents the Hubble distance.
Rather than computing the sound-horizon scale $r_d$ from early--Universe physics,
we treat it phenomenologically through the parameter combination $h\,r_d$,
thereby minimizing assumptions associated with recombination physics and baryon-density priors \cite{DESI:2024mwx,Colgain:2024xqj}.

\medskip
\noindent\textbf{(b) Type Ia supernovae:}
We employ three independent Type~Ia supernova compilations:
the Union3 binned dataset \cite{Scolnic:2021amr},
the Pantheon+SH0ES (PPS) sample \cite{Rubin:2023jdq},
and the DES--SN5YR Dovekie compilation \cite{DES:2024jxu,DES:2025sig}.
The Union3 sample spans the redshift range $0.01<z<2.26$,
while PPS covers $0.001<z<2.26$ including peculiar--velocity corrections.
For DES--SN5YR, we use the DES-Dovekie sample calibrated within the
SALT3.DES5YR framework.
This compilation contains 1820 supernovae distributed over
$0.01\lesssim z\lesssim1.13$,
including 1623 likely DES supernovae together with 197 low--redshift objects.
All supernova likelihoods are constructed using the full covariance matrices
provided by the corresponding collaborations, together with analytic
marginalization over the nuisance absolute magnitude parameter.

\medskip
\noindent\textbf{(c) Cosmic chronometers:}
Direct measurements of the Hubble parameter $H(z)$ are incorporated using the
cosmic chronometer method \cite{Gadbail:2024rpp}, which exploits the
differential age evolution of passively evolving galaxies.
The dataset consists of 32 independent measurements
\cite{Gaztanaga:2008xz,Oka:2013cba}
covering the redshift range $0.07<z<2.36$ and provides a direct probe of the
background expansion rate independent of any cosmological distance ladder.

\medskip
\noindent\textbf{(d) Growth-rate data:}
To constrain the evolution of matter perturbations, we additionally include
measurements of the growth-rate observable $f\sigma_8(z)$ obtained from DESI-DR2
and additional non-overlapping redshift-space distortion datasets
\cite{Sagredo:2018ahx,DESI:2024jxi}.
The compilation consists of 10 independent measurements spanning the range
$0.02<z<1.944$ and provides a direct probe of the growth history of
large--scale structures in the late--time Universe.

% ============================================================
\subsection{Model framework and growth formalism}
\label{subsec:GZ_background_growth}
% ============================================================

The cosmological evolution in the Gong--Zhang framework is governed by both the
background expansion history and the growth of matter perturbations.
In a spatially flat FLRW spacetime, the luminosity distance is given by
\cite{Hu:1995en}
\[
d_L(z)
=
(1+z)\int_0^z\frac{dx}{H(x)}
=
\frac{1+z}{H_0}\int_0^z\frac{dx}{E(x)},
\]
where the dimensionless Hubble parameter is defined as
$E(z)\equiv H(z)/H_0$.
The transverse comoving distance becomes
$D_M(z)=d_L(z)/(1+z)$,
while the volume--averaged BAO distance is
\[
D_V(z)
=
\left[
z\,D_M^2(z)\,D_H(z)
\right]^{1/3}.
\]

For the standard $\Lambda$CDM model, the normalized Hubble expansion reads
\begin{equation}
E^2(z)
=
\Omega_r(1+z)^4
+\Omega_m(1+z)^3
+1-\Omega_m-\Omega_r .
\end{equation}

For the Gong--Zhang Type~I (GZ1) parametrization,
\begin{equation}
w(z)=\frac{\omega_0}{1+z},
\end{equation}
the corresponding expansion history becomes
\begin{equation}
E^2(z)
=
\Omega_r(1+z)^4
+\Omega_m(1+z)^3
+\bigl(1-\Omega_m-\Omega_r\bigr)(1+z)^3
\exp\!\left(\frac{3\omega_0 z}{1+z}\right).
\end{equation}

For the Gong--Zhang Type~II (GZ2) parametrization,
\begin{equation}
w(z)=\frac{\omega_0 e^{z/(1+z)}}{1+z},
\end{equation}
the normalized Hubble expansion becomes
\begin{equation}
E^2(z)
=
\Omega_r(1+z)^4
+\Omega_m(1+z)^3
+\bigl(1-\Omega_m-\Omega_r\bigr)(1+z)^3
\exp\!\left[
3\omega_0\left(e^{\frac{z}{1+z}}-1\right)
\right].
\end{equation}

The impact of dynamical dark energy on structure formation is incorporated
through the linear growth factor $D(a)$, where the scale factor is related to
the redshift through $a=(1+z)^{-1}$.
The evolution of matter perturbations is computed using the
Linder--Jenkins growth equation \cite{Linder:2003dr},
\begin{equation}
D''(a)
+\frac{3}{2}
\left[
1-\frac{\omega(a)}{1+X(a)}
\right]
\frac{D'(a)}{a}
-\frac{3}{2}
\frac{X(a)}{1+X(a)}
\frac{D(a)}{a^2}
=0,
\end{equation}
where a prime denotes differentiation with respect to the scale factor $a$, and
\begin{equation}
X(a)=\frac{\Omega_m(a)}{1-\Omega_m(a)}.
\end{equation}
Throughout this analysis, dark-energy perturbations are neglected and the
evolution is studied within the sub-horizon linear regime.\\

This equation consistently accounts for the interplay between the evolving
dark--energy equation of state and the clustering of pressureless matter.
Together, the background expansion encoded in $E(a)$ and the growth factor
$D(a)$ fully determine the entropy evolution through the source term
$D(a)D'(a)$ appearing in the configuration--entropy equation, which will be
discussed in Sec.~\ref{sec:CE}. As a consequence, the Gong--Zhang parametrizations naturally generate distinct and observationally testable entropy--production histories. At the same time, their construction ensures compatibility with early--time cosmology, since dark--energy effects remain dynamically suppressed during the
radiation-- and matter--dominated eras.\\

From the linear growth factor $D(a)$, the logarithmic growth rate is defined as
\begin{equation}
f(a)
=
\frac{d\ln D}{d\ln a}
=
\frac{aD'(a)}{D(a)}.
\end{equation}

Since observational growth-rate measurements are conventionally reported as
functions of redshift, we subsequently express the growth observables in terms
of the redshift variable using the transformation $a=(1+z)^{-1}$.  The observable growth quantity therefore becomes,
becomes
\begin{equation}
f\sigma_8(z)
=
f(z)\,\sigma_8(z),
\end{equation}
with
\begin{equation}
\sigma_8(z)
=
\sigma_{8,0}\frac{D(z)}{D(0)}.
\end{equation}

To quantify departures from the standard cosmological model, we further define
the relative growth-rate deviation with respect to $\Lambda$CDM as
\begin{equation}
\Delta f(z)
=
\frac{
f_{\rm model}(z)-f_{\Lambda{\rm CDM}}(z)
}{
f_{\Lambda{\rm CDM}}(z)
}.
\end{equation}

The Gong--Zhang parametrizations therefore modify both the background expansion
history and the growth of matter perturbations while approximately preserving an
effective matter--dominated behaviour at early times due to the dynamical
suppression of dark energy at high redshift.

% ============================================================
\subsection{MCMC methodology and statistical analysis}
% ============================================================
We perform parameter estimation using Markov Chain Monte Carlo (MCMC)
simulations \cite{ForemanMackey:2013} implemented through the publicly
available \texttt{emcee} sampler \cite{Lewis:2019xzd}.
The total likelihood is constructed from the combined contributions of the
Type~Ia supernova, cosmic chronometer, BAO, and growth-rate datasets as
\begin{equation}
\mathcal{L}
=
\mathcal{L}_{\rm SN}\,
\mathcal{L}_{\rm OHD}\,
\mathcal{L}_{\rm BAO}\,
\mathcal{L}_{f\sigma_8},
\end{equation}
and sampled over the five--dimensional parameter space
\begin{equation}
\theta=
\{\Omega_m,\;h,\;h\,r_d,\;\omega_0,\;\sigma_8\}.
\end{equation}

Here, $\Omega_m$ denotes the present matter density parameter,
$h=H_0/100$ is the dimensionless Hubble constant,
$\omega_0$ characterizes the present value of the dark--energy equation of
state, and the sound-horizon scale $r_d$ is treated phenomenologically through
the parameter combination $h\,r_d$.
This setup enables a minimally model--dependent late--time analysis,
avoiding explicit assumptions associated with recombination physics,
baryon-density priors, or early--Universe calibration of the sound horizon.
The parameter $\sigma_8$ is also treated phenomenologically and constrained
directly through the observational $f\sigma_8(z)$ measurements, without
imposing priors from a full primordial CMB perturbation analysis.
The resulting posterior distributions and confidence contours are analyzed
using the \texttt{GetDist} package.\\

To quantitatively assess the statistical performance of the Gong--Zhang
parametrizations relative to the reference $\Lambda$CDM model,
we employ the Akaike Information Criterion (AIC) and Bayesian Information
Criterion (BIC), defined respectively as
\begin{equation}
\mathrm{AIC}
=
\chi^2_{\min}+2m,
\end{equation}
and
\begin{equation}
\mathrm{BIC}
=
\chi^2_{\min}+m\ln N,
\end{equation}
where $\chi^2_{\min}$ is the minimum chi--square value,
$m$ denotes the number of free parameters,
and $N$ is the total number of observational data points.

For each dataset combination, we compute
\[
\Delta\mathrm{AIC}
=
\mathrm{AIC}_{\rm model}
-
\mathrm{AIC}_{\Lambda{\rm CDM}},
\qquad
\Delta\mathrm{BIC}
=
\mathrm{BIC}_{\rm model}
-
\mathrm{BIC}_{\Lambda{\rm CDM}},
\]
such that negative values indicate a statistical preference for the
corresponding Gong--Zhang parametrization relative to $\Lambda$CDM
\cite{deCruzPerez:2024shj}.\\

{\color{darkblue}Following the standard information-criterion interpretation, the strength of the evidence is quantified by the magnitude $|\Delta\mathrm{IC}|$, while the sign of $\Delta\mathrm{IC}$ determines the preferred model. Specifically, $\Delta\mathrm{IC}<0$ indicates a preference for the corresponding dark-energy parametrization relative to $\Lambda$CDM, whereas $\Delta\mathrm{IC}>0$ implies that $\Lambda$CDM is preferred. In terms of the evidence strength, $|\Delta\mathrm{IC}|<2$ corresponds to weak or inconclusive evidence, $2\le|\Delta\mathrm{IC}|<6$ indicates positive evidence, $6\le|\Delta\mathrm{IC}|<10$ implies strong evidence, and $|\Delta\mathrm{IC}|\ge10$ denotes decisive evidence in favour of the model with the smaller information criterion value.}

% ===============================================================
\section{Results based on observational analysis}
\label{sec:obs}
\begin{figure}[h!]
    \centering
    \begin{minipage}[b]{0.48\textwidth}
        \centering
        \includegraphics[width=\textwidth]{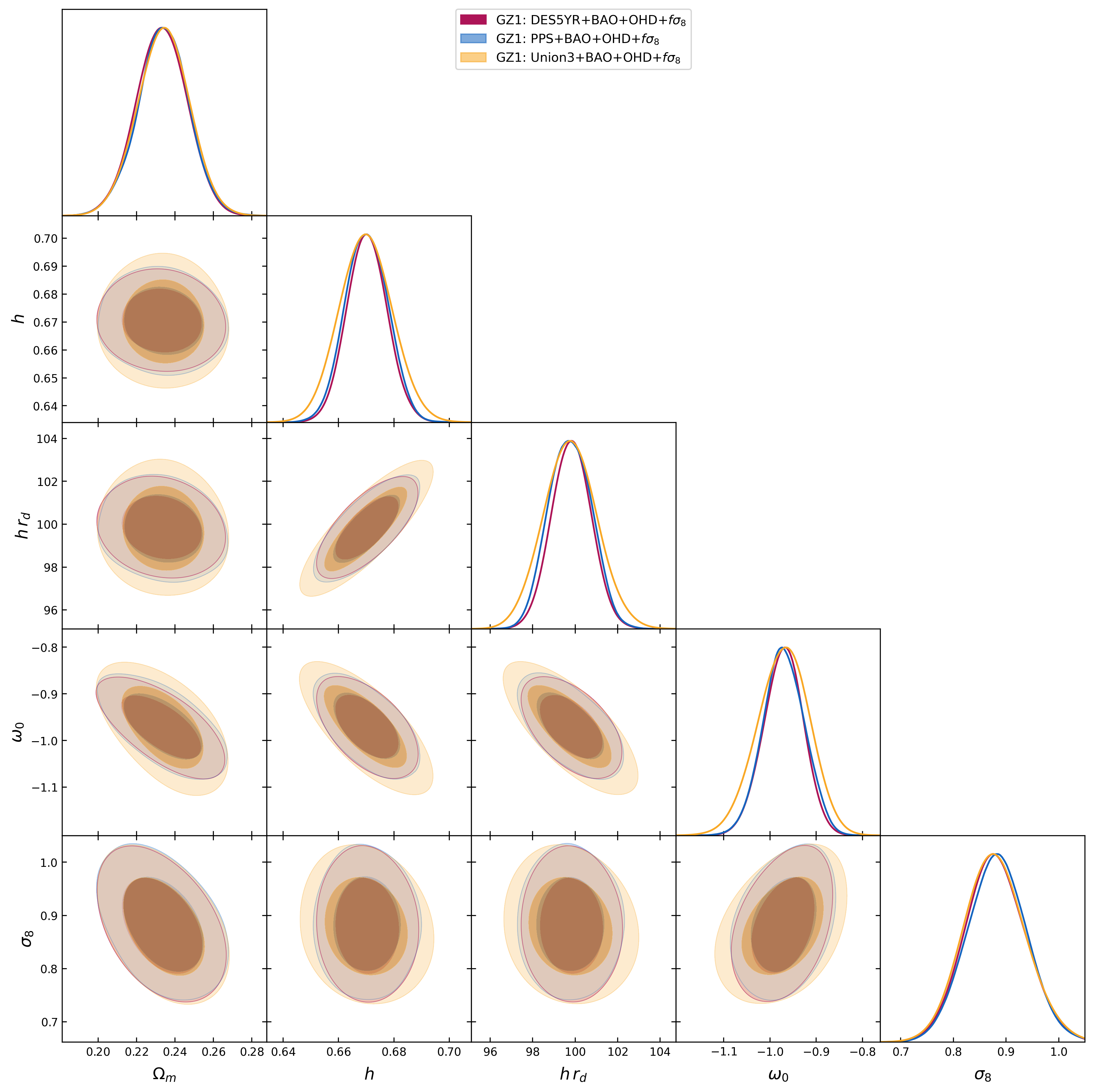}
%        \caption*{(a) 2D contours of $\Omega_m$--$\omega_0$}
    \end{minipage}
    \hfill
    \begin{minipage}[b]{0.48\textwidth}
        \centering
        \includegraphics[width=\textwidth]{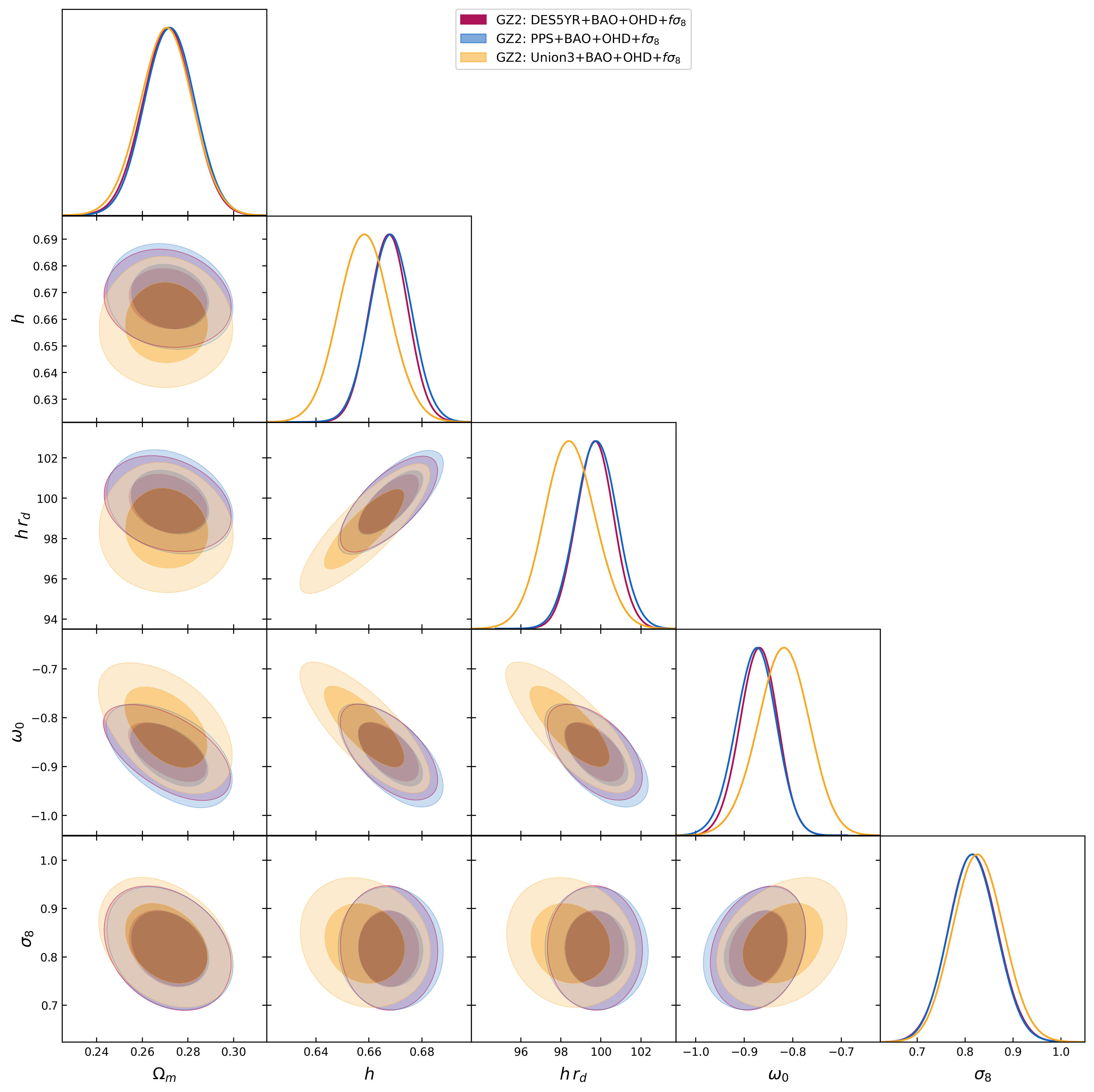}
%        \caption*{(b) 1D posteriors of $\omega_0$ in the GZ1 model}
    \end{minipage}

    \caption{Marginalized one-- and two--dimensional posterior distributions obtained from three SN+BAO+OHD +$f\sigma_8$ dataset combinations for the Gong--Zhang Type~I (left) and Type~II (right) models.}
    \label{fig:p-gz}
\end{figure}

% ===============================================================
 We constrain the Gong--Zhang Type~I (GZ1) and Type~II (GZ2) dark energy parametrizations using three independent combinations of late-time cosmological datasets,  U3+BAO+OHD+$f\sigma_8$, Pantheon+SH0ES+BAO+OHD+$f\sigma_8$, and DES--SN5YR+BAO+OHD+$f\sigma_8$. These combined probes simultaneously constrain the background expansion history and the growth of cosmic structures through Type~Ia supernova luminosity distances, cosmic chronometer measurements of $H(z)$, DESI DR2 BAO observables $D_M(z)/r_d$ and $D_H(z)/r_d$, together with the growth-rate observable $f\sigma_8(z)$. The corresponding marginalized one-- and two--dimensional posterior distributions are shown in Fig.~\ref{fig:p-gz}, where the left and right panels correspond to the GZ1 and GZ2 models, respectively.\\

For the GZ1 parametrization, the matter density parameter is constrained around $\Omega_m \sim 0.22$--$0.25$, while the Hubble parameter remains in the range $h \sim 0.66$--$0.69$. The parameter combination $h r_d$ is tightly constrained by the BAO measurements and exhibits a clear positive correlation with $h$. The equation-of-state parameter favors values close to the cosmological-constant regime, typically around $\omega_0 \sim -1$, with the contours displaying the expected anti-correlation between $\Omega_m$ and $\omega_0$. The inclusion of the $f\sigma_8$ growth data further constrains the clustering amplitude $\sigma_8$, which is found to lie near $\sigma_8 \sim 0.85$--$0.95$ with relatively small degeneracy with the remaining background parameters. Overall, the GZ1 model remains close to the standard $\Lambda$CDM scenario, exhibiting only mild departures from a cosmological-constant behaviour. The corresponding constraints for the GZ2 parametrization are shown in the right panel of Fig.~\ref{fig:p-gz}. In comparison with GZ1, the GZ2 model prefers slightly larger values of $\Omega_m$ and comparatively less negative values of the equation-of-state parameter, typically within the range $\omega_0 \sim -0.95$ to $-0.75$. The posterior contours involving $\omega_0$ are visibly tighter and less elongated, indicating a reduced degeneracy between the dark-energy dynamics and the background cosmological parameters. Similar to the GZ1 case, the BAO measurements strongly correlate $h$ and $h r_d$, while the growth data tightly constrain $\sigma_8$ around $\sigma_8 \sim 0.80$--$0.90$. The GZ2 parametrization also exhibits a comparatively moderate  correlation between $\omega_0$ and $\sigma_8$, suggesting a more direct connection between the dark-energy dynamics and the late-time growth of matter perturbations. Across all three dataset combinations, the DES--SN5YR+BAO+OHD+$f\sigma_8$ sample provides the tightest constraints for both parametrizations. Overall, the inclusion of the growth-rate data significantly reduces the allowed parameter space and demonstrates that both Gong--Zhang models remain compatible with the observed late-time expansion and structure-growth histories, with GZ2 exhibiting a comparatively less degenerate and more tightly constrained parameter space.\\

\subsection{Correlation Analysis}
\label{sec:correlation}

\begin{figure}[h!]
    \centering

    % ===================== GZ1 =====================
%    \textbf{}\par\medskip
    \begin{minipage}[b]{0.32\textwidth}
        \centering
        \includegraphics[width=\textwidth]{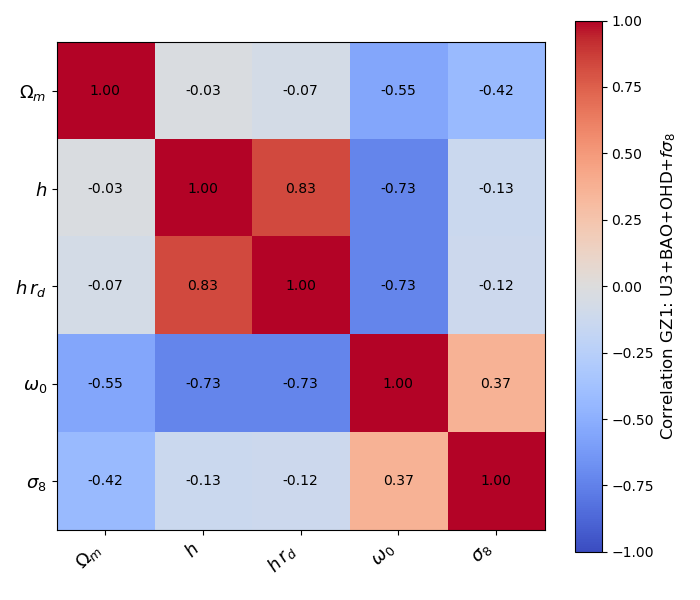}
        \caption*{(a) GZ1: Union3 + BAO + OHD + f$\sigma_8$}
    \end{minipage}
    \hfill
    \begin{minipage}[b]{0.32\textwidth}
        \centering
        \includegraphics[width=\textwidth]{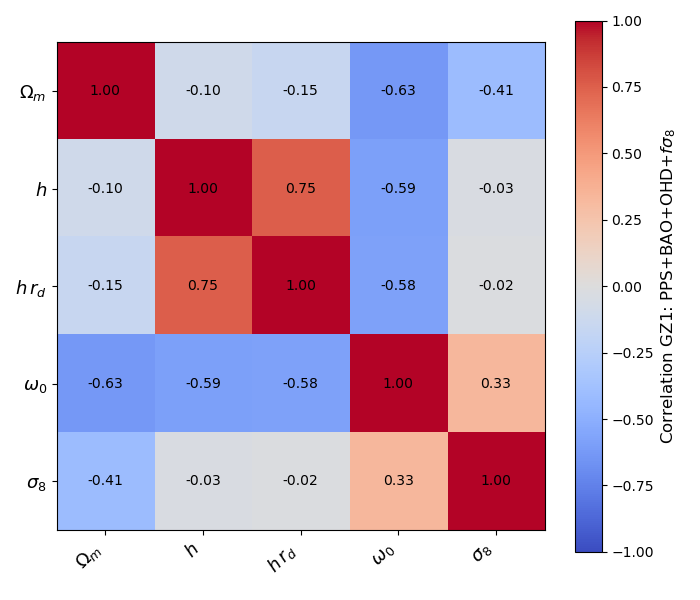}
        \caption*{(b) GZ1: Pantheon+SH0ES + BAO + OHD + f$\sigma_8$}
    \end{minipage}
    \hfill
    \begin{minipage}[b]{0.32\textwidth}
        \centering
        \includegraphics[width=\textwidth]{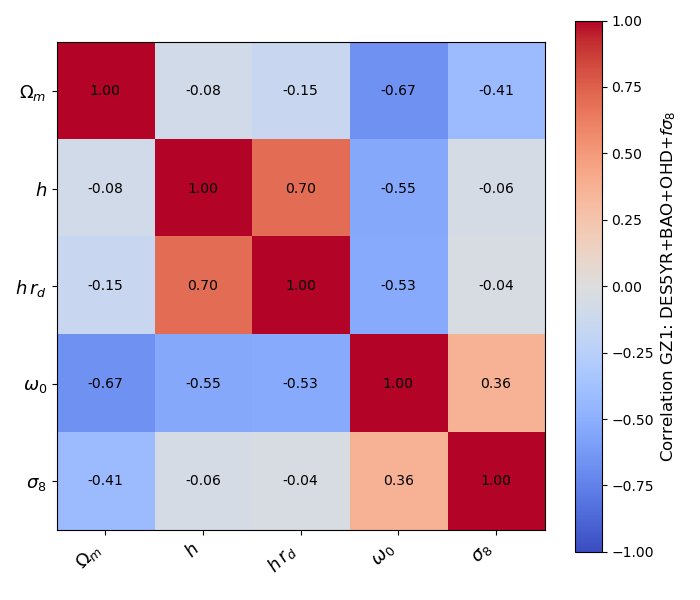}
        \caption*{(c) GZ1: DES--SN5YR + BAO + OHD + f$\sigma_8$}
    \end{minipage}

    \vspace{1em}

    % ===================== GZ2 =====================
%    \textbf{GZ2 Model}\par\medskip
    \begin{minipage}[b]{0.32\textwidth}
        \centering
        \includegraphics[width=\textwidth]{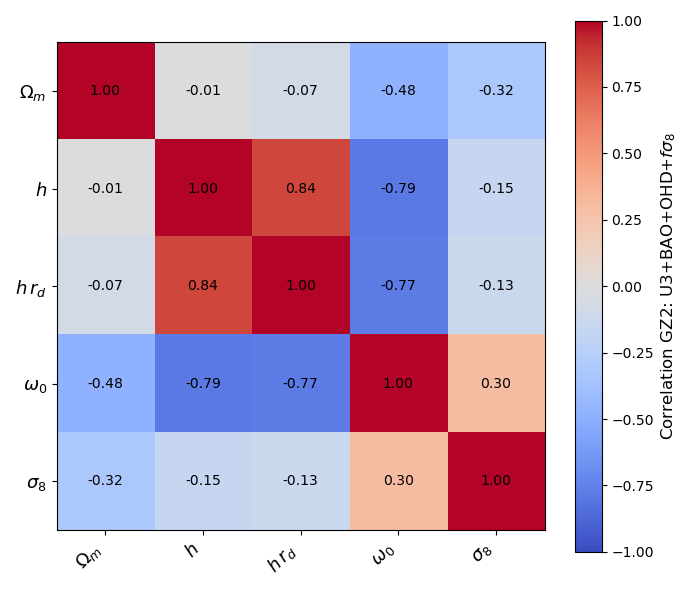}
        \caption*{(d) GZ2: Union3 + BAO + OHD + f$\sigma_8$}
    \end{minipage}
    \hfill
    \begin{minipage}[b]{0.32\textwidth}
        \centering
        \includegraphics[width=\textwidth]{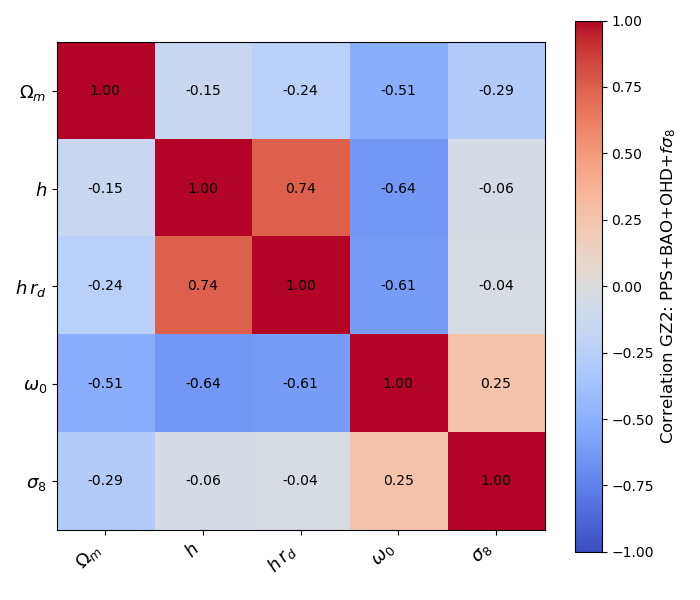}
        \caption*{(e) GZ2: Pantheon+SH0ES + BAO + OHD + f$\sigma_8$}
    \end{minipage}
    \hfill
    \begin{minipage}[b]{0.32\textwidth}
        \centering
        \includegraphics[width=\textwidth]{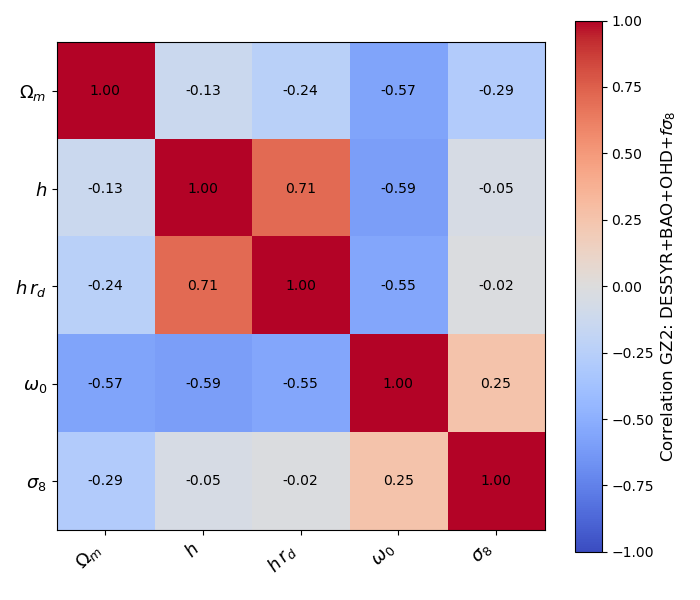}
        \caption*{(f) GZ2: DES--SN5YR + BAO + OHD + f$\sigma_8$}
    \end{minipage}

    \caption{Correlation matrices for the Gong--Zhang Type~I (top row) and
    Type~II (bottom row) models obtained from three SN+BAO+OHD+ f$\sigma_8$ datasets
    combinations.}
    \label{fig:GZ-crs}
\end{figure}

 Fig.~\ref{fig:GZ-crs} presents the correlation matrices for the Gong--Zhang Type~I (top row) and Type~II (bottom row) parametrizations obtained from the three combined late-time datasets: U3+BAO+OHD+$f\sigma_8$, Pantheon+SH0ES+BAO+OHD+$f\sigma_8$, and DES--SN5YR+BAO+OHD+$f\sigma_8$. The matrices illustrate the mutual correlations among the cosmological parameters ${\Omega_m, h, h r_d, \omega_0, \sigma_8}$ and reveal the underlying degeneracy structure of each model. For both GZ1 and GZ2, a strong positive correlation is consistently observed between $h$ and $h r_d$, mainly driven by the BAO measurements that tightly constrain the cosmic distance scale. At the same time, the matter density parameter $\Omega_m$ shows a pronounced anti--correlation with the dark-energy equation-of-state parameter $\omega_0$, indicating that an increase in the matter content can be partially compensated by a more negative dark-energy behaviour. The GZ1 model generally exhibits comparatively stronger parameter degeneracies, particularly among $\Omega_m$, $h$, $h r_d$, and $\omega_0$, leading to more elongated posterior structures. In contrast, the GZ2 parametrization shows relatively weaker correlations involving $\omega_0$, especially in the Pantheon+SH0ES and DES--SN5YR combinations, suggesting a more effective separation between the dark-energy dynamics and the background expansion parameters. The correlations involving the clustering amplitude $\sigma_8$ remain moderate in both models, with a mild positive correlation with $\omega_0$ and a weak anti--correlation with $\Omega_m$. Overall, the reduced parameter degeneracies observed in GZ2 are consistent with the tighter posterior contours obtained in the MCMC analysis and indicate that the GZ2 parametrization provides a comparatively less degenerate description of late-time cosmic acceleration within the observational datasets considered here.

\subsection{Analysis from Jeffreys Plot comparison}
\label{sec:Jeffrey}

\begin{figure}[h!]
    \centering
        \includegraphics[width=0.8\textwidth]{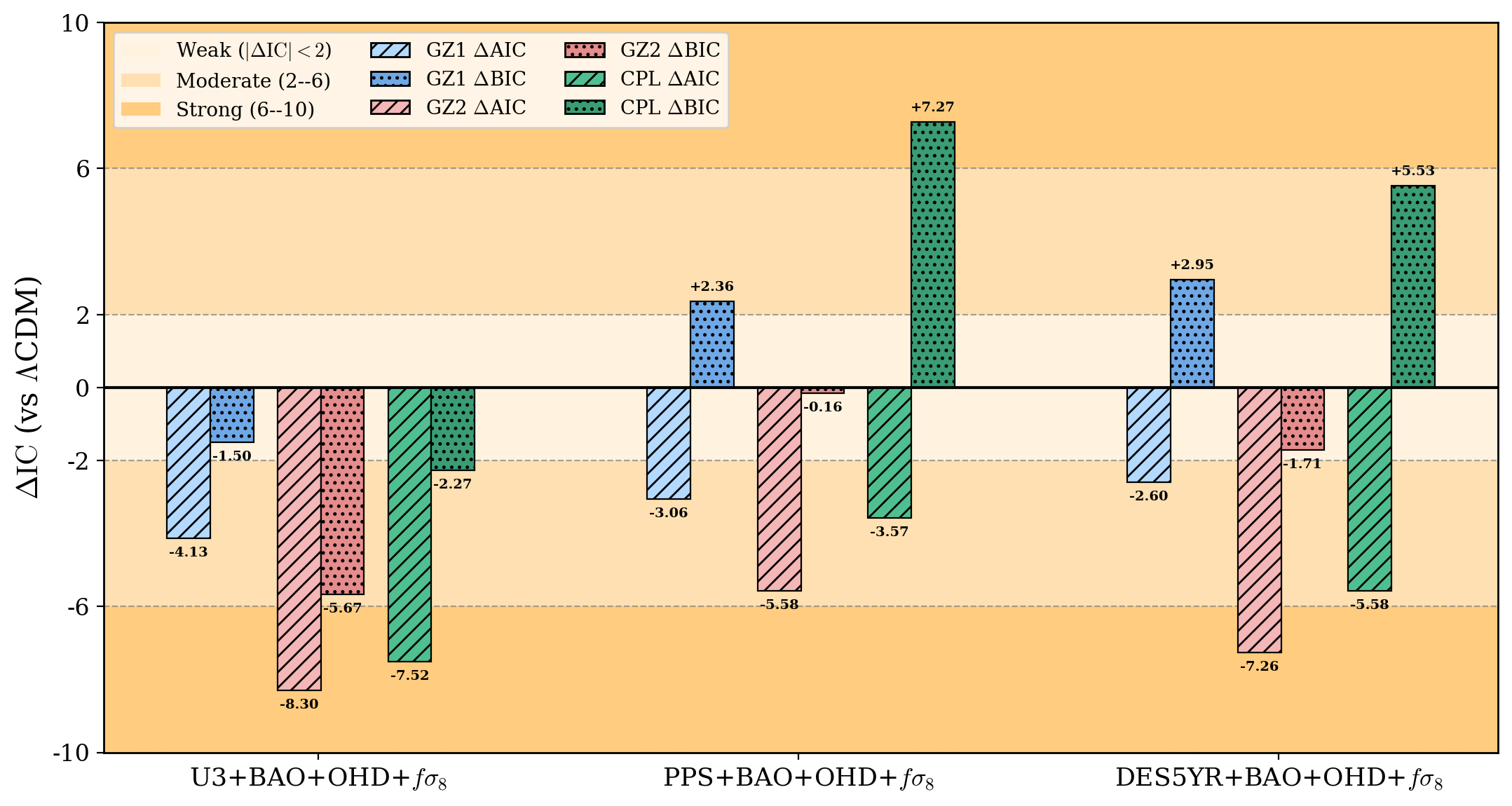}
            \caption{{\color{darkblue}Jeffreys--scale comparison of the GZ1, GZ2, and CPL parametrizations relative to $\Lambda$CDM using the signed quantities $\Delta{\rm AIC}$ and $\Delta{\rm BIC}$. Negative (positive) values indicate a preference for the corresponding parametrization ($\Lambda$CDM). The shaded regions represent the Jeffreys-scale strength based on $|\Delta{\rm IC}|$.}}
    \label{fig:GZ-Jeffreys}
\end{figure}

{\color{darkblue} Fig.~\ref{fig:GZ-Jeffreys} presents the Jeffreys--scale comparison of the Gong--Zhang Type~I (GZ1), Gong--Zhang Type~II (GZ2), and CPL\footnote{  
The Chevallier--Polarski--Linder (CPL) parametrization
\cite{Chevallier:2000qy} is characterized by the equation of state
$\omega(z)=\omega_0+\omega_a\frac{z}{1+z}$,
where $\omega_0$ represents the present value of the dark--energy equation of
state and $\omega_a$ quantifies its dynamical evolution with redshift.
For a spatially flat Universe, the corresponding normalized Hubble expansion
rate is given by
\[
E^2(z)=
\Omega_m(1+z)^3
+\Omega_r(1+z)^4
+\left(1-\Omega_m-\Omega_r\right)
(1+z)^{3(1+\omega_0+\omega_a)}
\exp\!\left[-\frac{3\omega_a z}{1+z}\right].
\]
For more details, see Appendix~\ref{app:CPL}.
} dark energy parametrizations relative to the reference $\Lambda$CDM cosmology using the combined late--time datasets Union3+BAO+OHD+$f\sigma_8$, Pantheon+SH0ES+BAO+OHD+$f\sigma_8$, and DES--SN5YR+BAO+OHD+$f\sigma_8$. The comparison is based on the Akaike and Bayesian information criteria through the quantities $\Delta{\rm AIC}$ and $\Delta{\rm BIC}$, which quantify the relative statistical performance of each parametrization with respect to $\Lambda$CDM. The corresponding marginalized posterior distributions for the CPL parametrization \cite{Chevallier:2000qy} are presented separately in Appendix.~\ref{app:CPL}.\\

For the Union3+BAO+OHD+$f\sigma_8$ dataset combination, all three parametrizations yield negative values of both $\Delta{\rm AIC}$ and $\Delta{\rm BIC}$, indicating a statistical preference relative to $\Lambda$CDM according to both criteria. The GZ2 parametrization exhibits the largest negative values, with $\Delta{\rm AIC}=-8.30$ and $\Delta{\rm BIC}=-5.67$, while the CPL model gives $\Delta{\rm AIC}=-7.52$ and $\Delta{\rm BIC}=-2.27$. The GZ1 parametrization also yields negative values of both criteria, although with comparatively smaller magnitudes. For the PPS+BAO+OHD+$f\sigma_8$ dataset combination, all three parametrizations continue to produce negative $\Delta{\rm AIC}$ values. However, the corresponding $\Delta{\rm BIC}$ values show a more mixed behaviour. In particular, GZ1 and CPL yield positive values, $\Delta{\rm BIC}=+2.36$ and $+7.27$, respectively, indicating a preference for $\Lambda$CDM once the additional model complexity is taken into account. For GZ2, the BIC difference remains close to zero, with $\Delta{\rm BIC}=-0.16$, suggesting that the model is statistically comparable to $\Lambda$CDM according to this criterion. A similar pattern is observed for the DES--SN5YR+BAO+OHD+$f\sigma_8$ dataset combination. While all three parametrizations have negative $\Delta{\rm AIC}$ values, the BIC values for GZ1 and CPL become positive, with $\Delta{\rm BIC}=+2.95$ and $+5.53$, respectively, implying a preference for $\Lambda$CDM. In contrast, GZ2 retains negative values for both criteria, with $\Delta{\rm AIC}=-7.26$ and $\Delta{\rm BIC}=-1.71$.\\

Overall, the information--criteria analysis indicates that the conclusions depend on both the sign and magnitude of the information--criterion differences. Although all three parametrizations generally provide lower AIC values than $\Lambda$CDM, the stronger complexity penalty incorporated in BIC weakens or removes this preference for GZ1 and CPL in some dataset combinations. Among the models considered, GZ2 exhibits the most consistent behaviour across the datasets, maintaining negative values of both $\Delta{\rm AIC}$ and $\Delta{\rm BIC}$. These results are broadly consistent with the best--fit cosmological parameters summarized in Table~\ref{tab:tab1}.}

\begin{table*}
\centering
\scriptsize
\setlength{\tabcolsep}{2.1pt}
\renewcommand{\arraystretch}{1.15}

\resizebox{\textwidth}{!}{%

\begin{tabular}{llccccccccc}

\hline\hline

Dataset & Model
& $\Omega_{m0}$
& $h$
& $h r_d$
& $\omega_0$
& $\omega_a$
& $\sigma_8$
& AIC
& BIC
& $\Delta$AIC / $\Delta$BIC \\

\hline

% ============================================================
% UNION3
% ============================================================

\multirow{4}{*}{Union3+BAO+OHD+$f\sigma_8$}

& $\Lambda$CDM
& $0.2982^{+0.009}_{-0.009}$
& $0.6821^{+0.006}_{-0.006}$
& $101.60^{+0.89}_{-0.89}$
& --
& --
& $0.7688^{+0.046}_{-0.046}$
& 90.77
& 101.27
& $0.00 / 0.00$ \\

& GZ1
& $0.2341^{+0.013}_{-0.013}$
& $0.6699^{+0.009}_{-0.009}$
& $99.78^{+1.25}_{-1.25}$
& $-0.9711^{+0.056}_{-0.056}$
& --
& $0.8792^{+0.059}_{-0.059}$
& 86.64
& 99.76
& $-4.13 / -1.50$ \\

& GZ2
& $0.2705^{+0.011}_{-0.011}$
& $0.6584^{+0.009}_{-0.009}$
& $98.47^{+1.27}_{-1.27}$
& $-0.8195^{+0.053}_{-0.053}$
& --
& $0.8280^{+0.053}_{-0.053}$
& 82.47
& 95.59
& $-8.30 / -5.67$ \\

& CPL
& $0.3063^{+0.015}_{-0.015}$
& $0.6546^{+0.01}_{-0.01}$
& $98.24^{+1.29}_{-1.29}$
& $-0.7461^{+0.089}_{-0.089}$
& $-0.5638^{+0.48}_{-0.48}$
& $0.7821^{+0.049}_{-0.049}$
& 83.24
& 98.99
& $-7.52 / -2.27$ \\

\hline

% ============================================================
% PPS
% ============================================================

\multirow{4}{*}{PPS+BAO+OHD+$f\sigma_8$}

& $\Lambda$CDM
& $0.2991^{+0.009}_{-0.009}$
& $0.6815^{+0.006}_{-0.006}$
& $101.51^{+0.85}_{-0.85}$
& --
& --
& $0.7691^{+0.046}_{-0.046}$
& 1474.54
& 1496.23
& $0.00 / 0.00$ \\

& GZ1
& $0.2339^{+0.013}_{-0.013}$
& $0.6702^{+0.008}_{-0.008}$
& $99.78^{+1.00}_{-1.00}$
& $-0.9697^{+0.045}_{-0.045}$
& --
& $0.8835^{+0.057}_{-0.057}$
& 1471.48
& 1498.58
& $-3.06 / +2.36$ \\

& GZ2
& $0.2721^{+0.011}_{-0.011}$
& $0.6683^{+0.008}_{-0.008}$
& $99.79^{+1.02}_{-1.02}$
& $-0.8758^{+0.042}_{-0.042}$
& --
& $0.8165^{+0.05}_{-0.05}$
& 1468.97
& 1496.07
& $-5.58 / -0.16$ \\

& CPL
& $0.2931^{+0.014}_{-0.014}$
& $0.6685^{+0.008}_{-0.008}$
& $99.97^{+1.01}_{-1.01}$
& $-0.8723^{+0.056}_{-0.056}$
& $-0.1572^{+0.366}_{-0.366}$
& $0.7855^{+0.05}_{-0.05}$
& 1470.97
& 1503.49
& $-3.57 / +7.27$ \\

\hline

% ============================================================
% DES5YR
% ============================================================

\multirow{4}{*}{DES5YR+BAO+OHD+$f\sigma_8$}

& $\Lambda$CDM
& $0.3021^{+0.008}_{-0.008}$
& $0.6799^{+0.006}_{-0.006}$
& $101.27^{+0.82}_{-0.82}$
& --
& --
& $0.7653^{+0.046}_{-0.046}$
& 1703.49
& 1725.69
& $0.00 / 0.00$ \\

& GZ1
& $0.2331^{+0.013}_{-0.013}$
& $0.6704^{+0.007}_{-0.007}$
& $99.83^{+0.94}_{-0.94}$
& $-0.9709^{+0.043}_{-0.043}$
& --
& $0.8803^{+0.057}_{-0.057}$
& 1700.89
& 1728.64
& $-2.60 / +2.95$ \\

& GZ2
& $0.2712^{+0.011}_{-0.011}$
& $0.6676^{+0.007}_{-0.007}$
& $99.71^{+0.93}_{-0.93}$
& $-0.8701^{+0.039}_{-0.039}$
& --
& $0.8171^{+0.051}_{-0.051}$
& 1696.23
& 1723.98
& $-7.26 / -1.71$ \\

& CPL
& $0.2988^{+0.014}_{-0.014}$
& $0.6672^{+0.007}_{-0.007}$
& $99.86^{+0.93}_{-0.93}$
& $-0.8345^{+0.065}_{-0.065}$
& $-0.4035^{+0.428}_{-0.428}$
& $0.7796^{+0.049}_{-0.049}$
& 1697.92
& 1731.22
& $-5.58 / +5.53$ \\

\hline\hline

\end{tabular}%

}

\caption{
Best--fit cosmological parameters and information criteria for
$\Lambda$CDM, GZ1, GZ2, and CPL models using the
Union3/PPS/DES5YR + BAO + OHD + $f\sigma_8$ datasets.
Errors correspond to the $1\sigma$ confidence interval.
All $\Delta$AIC and $\Delta$BIC values are computed with respect to the
corresponding $\Lambda$CDM fit for each dataset.
}

\label{tab:tab1}

\end{table*}

\subsection{One and Two Dimensional Posterior Constraints}
\label{sec:posteriors}

\begin{figure}[h!]
\centering
\setlength{\tabcolsep}{2pt}

\begin{tabular}{cc}
\includegraphics[width=0.46\linewidth]{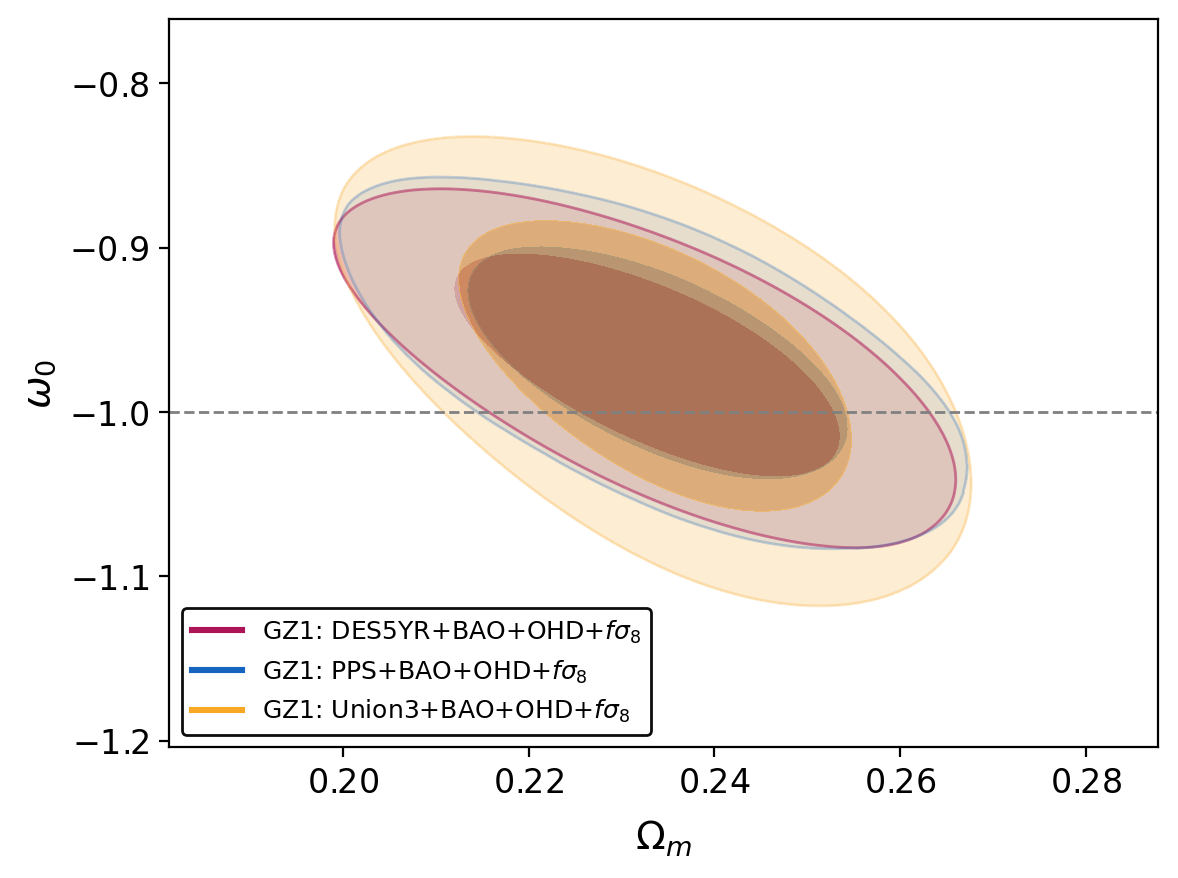} &
\includegraphics[width=0.46\linewidth]{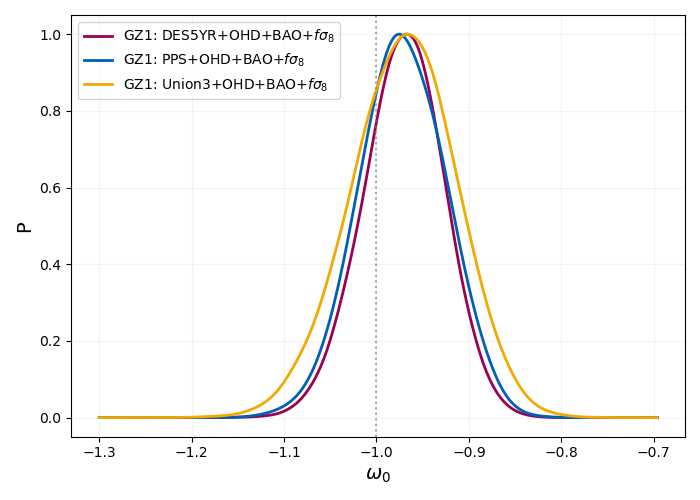} \\[-2pt]
{\scriptsize (a) GZ1: $\Omega_m$--$\omega_0$} &
{\scriptsize (b) GZ1: $\omega_0$ posterior} \\[4pt]

\includegraphics[width=0.46\linewidth]{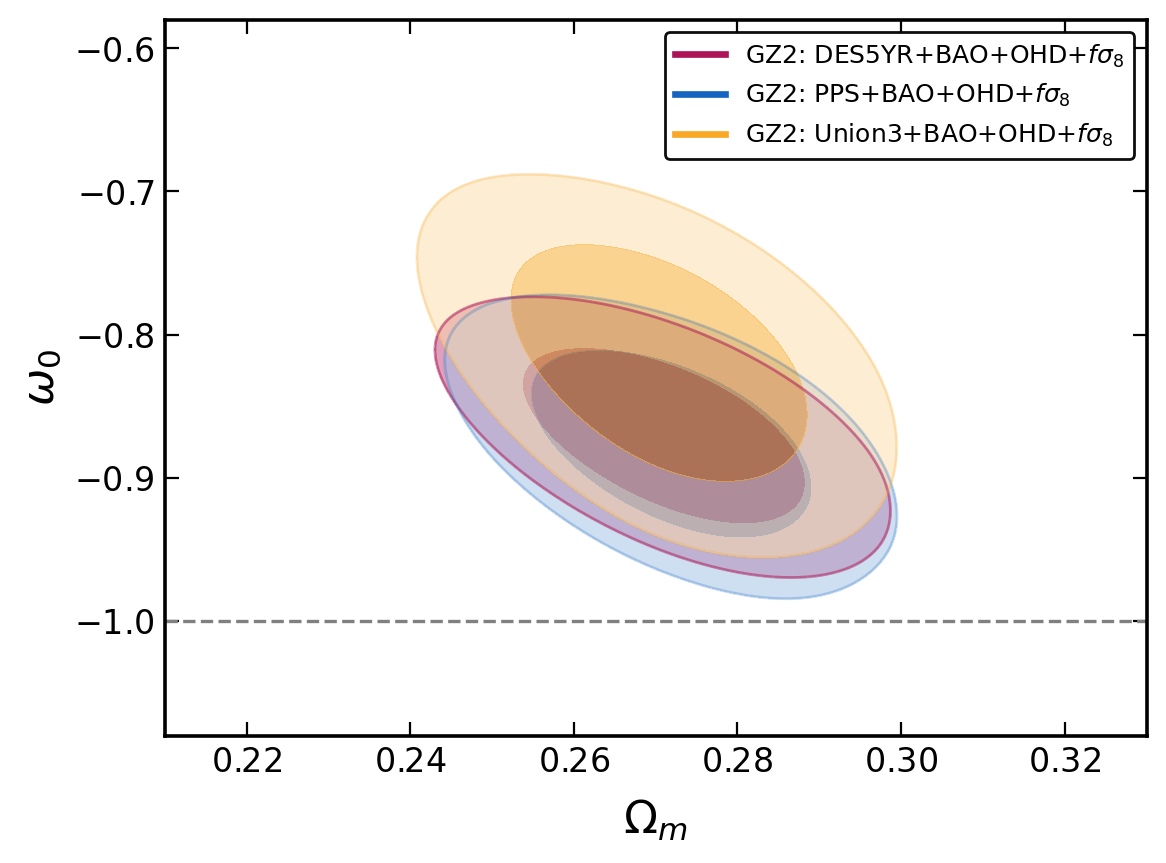} &
\includegraphics[width=0.46\linewidth]{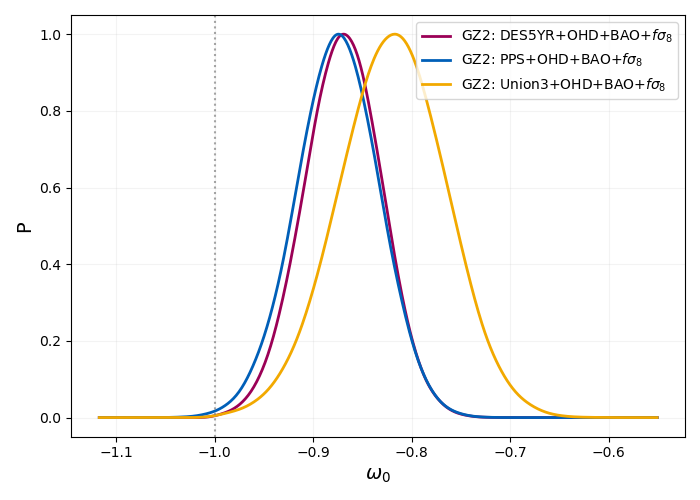} \\[-2pt]
{\scriptsize (c) GZ2: $\Omega_m$--$\omega_0$} &
{\scriptsize (d) GZ2: $\omega_0$ posterior}
\end{tabular}

\vspace{-0.5em}
\caption{ Marginalized constraints on the Gong--Zhang dark--energy models.
Panels (a,c) show the $68\%$ and $95\%$ confidence contours in the
$\Omega_m$--$\omega_0$ plane, while (b,d) display the corresponding one--dimensional
posteriors of $\omega_0$ for different SN+BAO+OHD+f$\sigma_8$ combinations.}
\label{fig:GZ_w0}
\end{figure}

Figure~\ref{fig:GZ_w0} presents the marginalized one-- and two--dimensional posterior distributions for the Gong--Zhang Type~I (GZ1) and Type~II (GZ2) dark--energy parametrizations obtained from the combined analyses of supernovae (DES--SN5YR, Pantheon+SH0ES, and Union3), BAO, cosmic chronometer $H(z)$, and $f\sigma_8$ observations. The upper panels correspond to the GZ1 model, where panel~(a) shows the $68\%$ and $95\%$ confidence contours in the $\Omega_m$--$\omega_0$ plane and panel~(b) displays the corresponding marginalized posterior distributions of $\omega_0$. All dataset combinations consistently favor values of $\omega_0$ close to the cosmological--constant limit $\omega_0=-1$, with only mild shifts among the different supernova compilations. The strong overlap of the confidence contours further indicates a high level of consistency between the observational datasets. The lower panels show the corresponding results for the GZ2 parametrization. Similar to GZ1, the confidence contours in the $\Omega_m$--$\omega_0$ plane exhibit substantial overlap among the different dataset combinations, demonstrating stable parameter estimation across the low--redshift probes considered. However, compared to GZ1, the GZ2 model prefers comparatively less negative values of $\omega_0$ together with slightly higher values of $\Omega_m$. The one--dimensional posterior distributions in panel~(d) are also noticeably narrower, indicating tighter constraints on $\omega_0$ and a comparatively reduced degeneracy between $\Omega_m$ and the dark--energy equation--of--state parameter. Overall, both parametrizations produce well--behaved and statistically consistent posterior distributions without showing any significant tension among the observational datasets.\\

\begin{figure}[h!]
    \centering
    
    \includegraphics[width=0.48\textwidth]{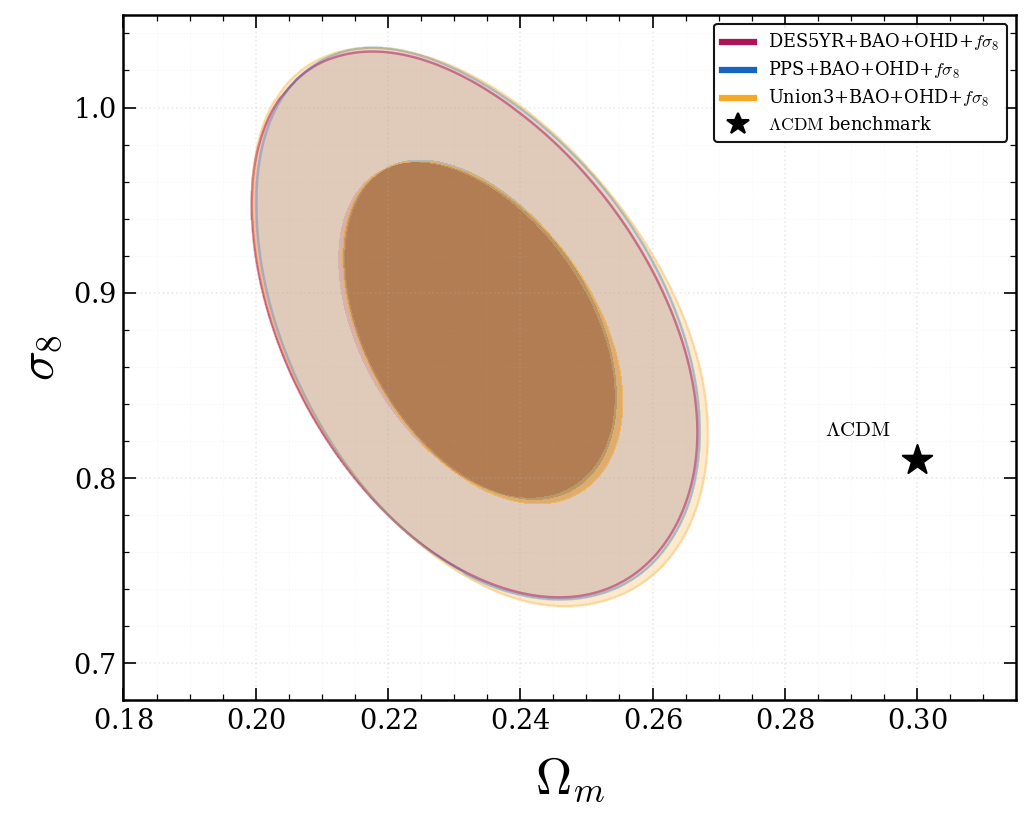}
    \hfill
    \includegraphics[width=0.48\textwidth]{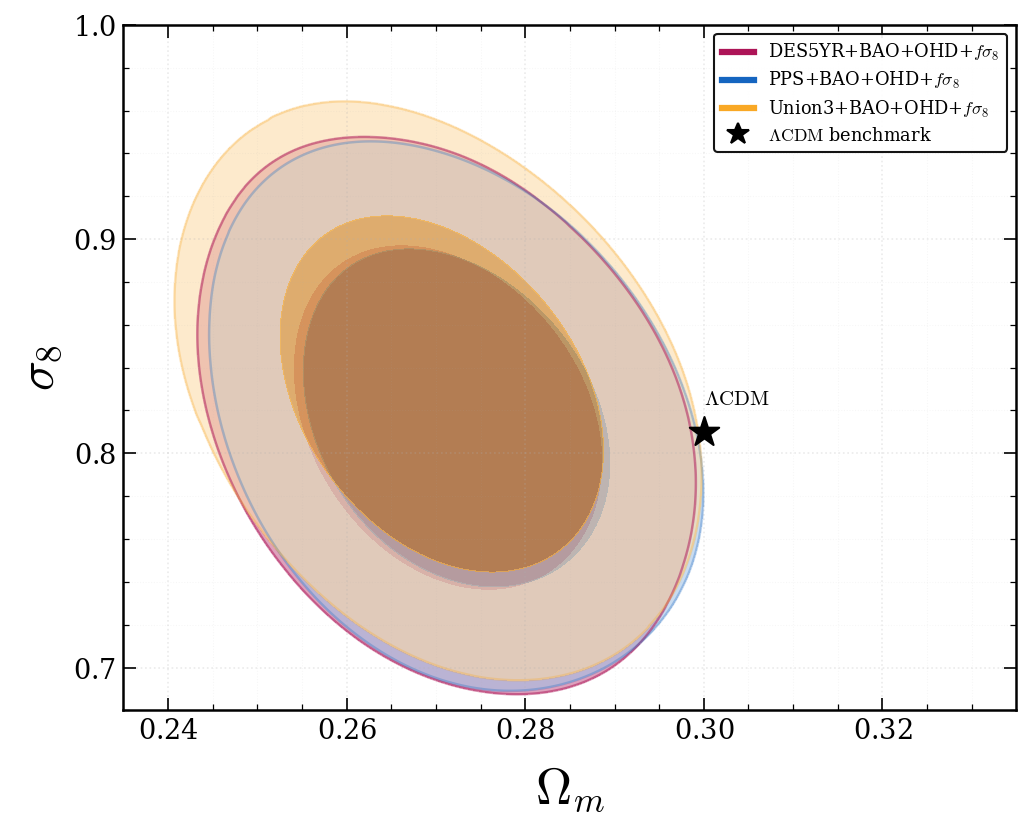}

    \caption{Marginalized two--dimensional posterior distributions of $\Omega_m-\sigma_8$ obtained from three SN+BAO+OHD+ $f\sigma_8$ dataset combinations for the Gong--Zhang Type~I (left) and Type~II (right) models and black star denotes the reference $\Lambda$CDM benchmark point.}
    \label{fig:s8-gz}
\end{figure}

Figure~\ref{fig:s8-gz} presents the marginalized constraints in the $\Omega_m$--$\sigma_8$ plane for the Gong--Zhang Type~I (GZ1) and Type~II (GZ2) parametrizations obtained from the combined SN+BAO+OHD+$f\sigma_8$ datasets. In both models, the inclusion of the observational growth-rate data leads to stable and well-constrained posterior regions with substantial overlap among the Union3, Pantheon+SH0ES, and DES5YR dataset combinations, indicating good consistency between the background and structure-growth probes. A clear anti-correlation between $\Omega_m$ and $\sigma_8$ is observed, reflecting the expected interplay between the present matter density and the amplitude of matter clustering. Compared to GZ1, the GZ2 parametrization yields slightly tighter and more localized contours in the $\Omega_m$--$\sigma_8$ plane, suggesting a comparatively reduced degeneracy in the growth sector and a reduced clustering-sector degeneracy. The proximity of the contours to the $\Lambda$CDM reference point further indicates that both Gong--Zhang parametrizations remain compatible with the observed late--time structure formation history while allowing mild departures from the standard cosmological scenario.

%
% 
%
%===============================================================
\subsection{Cosmographic and evolutionary analysis of the Gong–Zhang Models}
\label{subsec:GZ_cosmography}
%===============================================================

Cosmography provides a purely kinematical description of the cosmic expansion
history, independent of the underlying gravitational dynamics or the specific
matter content of the Universe.
It relies on a Taylor expansion of the scale factor around the present epoch
and therefore offers a model--independent framework to characterize late--time
cosmic acceleration and possible deviations from the standard $\Lambda$CDM
scenario.
The expansion history is encoded in the hierarchy of cosmographic parameters,
defined through successive time derivatives of the scale factor as
\begin{equation}
   H = \frac{\dot a}{a}, \qquad
   q = -\frac{\ddot a}{aH^2}, \qquad
   j = \frac{\dddot a}{aH^3}, \qquad
   s = \frac{a^{(4)}}{aH^4},
\end{equation}
where $H$ is the Hubble parameter, $q$ the deceleration parameter, $j$ the
jerk, and $s$ the snap.

For a general Friedmann expansion history expressed in terms of the normalized
Hubble parameter $E(z)=H(z)/H_0$, the cosmographic parameters can be written as
functions of redshift according to
\begin{align}
q(z)
  &= -1 + (1+z)\frac{E'(z)}{E(z)}, \\[1mm]
j(z)
  &= q(z)\bigl[2q(z)+1\bigr] + (1+z)\,q'(z), \\[1mm]
s(z)
  &= j(z)\bigl[3q(z)+2\bigr] + (1+z)\,j'(z),
\end{align}
where a prime denotes differentiation with respect to $z$.
The present--day values are defined as
$q_0=q(0)$, $j_0=j(0)$, and $s_0=s(0)$.
These quantities provide a transparent and physically intuitive way to compare
different dark--energy parametrizations on purely kinematical grounds.

We now apply this formalism to the Gong--Zhang dark--energy models.
For the Gong--Zhang Type~I (GZ1) parametrization,
\begin{equation}
w(z)=\frac{\omega_0}{1+z},
\end{equation}

For the Gong--Zhang Type~II (GZ2) parametrization,
\begin{equation}
w(z)=\frac{\omega_0 e^{z/(1+z)}}{1+z},
\end{equation}
the dark--energy contribution follows from the corresponding integral of the
equation of state.
In both cases, the cosmographic parameters are obtained from derivatives of
$E(z)$ evaluated at $z=0$.

At leading order, the two parametrizations yield an identical expression for
the present--day deceleration parameter,
\begin{equation}
q_0=\frac{1}{2}\Omega_m+\Omega_r
+\frac{1}{2}(1-\Omega_m-\Omega_r)(1+3\omega_0),
\end{equation}
ensuring comparable late--time acceleration. Although $\Omega_r$ is negligible at late times, it is retained for completeness.
Differences between the GZ1 and GZ2 models arise primarily in the higher--order
cosmographic parameters $j_0$ and $s_0$, which are sensitive to the detailed
redshift evolution of the dark--energy equation of state.
%

% ============================================================
% RED COLORED COSMOGRAPHY TABLE
% ============================================================

\begin{table*}[h!]
\centering
\scriptsize
\renewcommand{\arraystretch}{1.2}
\setlength{\tabcolsep}{6pt}

\begin{tabular}{lcccccc}
\hline\hline

& \multicolumn{2}{c}{Union3+BAO+OHD+$f\sigma_8$}
& \multicolumn{2}{c}{PPS+BAO+OHD+$f\sigma_8$}
& \multicolumn{2}{c}{DES5YR+BAO+OHD+$f\sigma_8$} \\

\cline{2-7}

& GZ1 & GZ2 & GZ1 & GZ2 & GZ1 & GZ2 \\

\hline

$q_{0}$

& $-0.6160^{+0.056}_{-0.056}$
& $-0.3964^{+0.053}_{-0.053}$

& $-0.6126^{+0.042}_{-0.042}$
& $-0.4546^{+0.04}_{-0.04}$

& $-0.6138^{+0.039}_{-0.039}$
& $-0.4501^{+0.036}_{-0.036}$

\\[3pt]

$j_{0}$

& $2.0308^{+0.24}_{-0.24}$
& $1.4198^{+0.17}_{-0.17}$

& $2.0129^{+0.183}_{-0.183}$
& $1.6001^{+0.142}_{-0.142}$

& $2.0133^{+0.177}_{-0.177}$
& $1.5814^{+0.131}_{-0.131}$

\\[3pt]

$s_{0}$

& $-5.2803^{+0.52}_{-0.52}$
& $-4.0456^{+0.485}_{-0.485}$

& $-5.2539^{+0.423}_{-0.423}$
& $-4.5456^{+0.397}_{-0.397}$

& $-5.2483^{+0.41}_{-0.41}$
& $-4.4929^{+0.371}_{-0.371}$

\\

\hline\hline

\end{tabular}

\caption{
Cosmographic parameters for GZ--Type~I (GZ1) and
GZ--Type~II (GZ2) obtained from the MCMC chains using
Union3/PPS/DES5YR + BAO + OHD + $f\sigma_8$ datasets.
Quoted values correspond to the mean $\pm 1\sigma$ uncertainty.
}

\label{tab:tab2}

\end{table*}

\begin{figure}[h!]
    \centering
    \begin{minipage}[b]{0.48\textwidth}
        \centering
        \includegraphics[width=\textwidth]{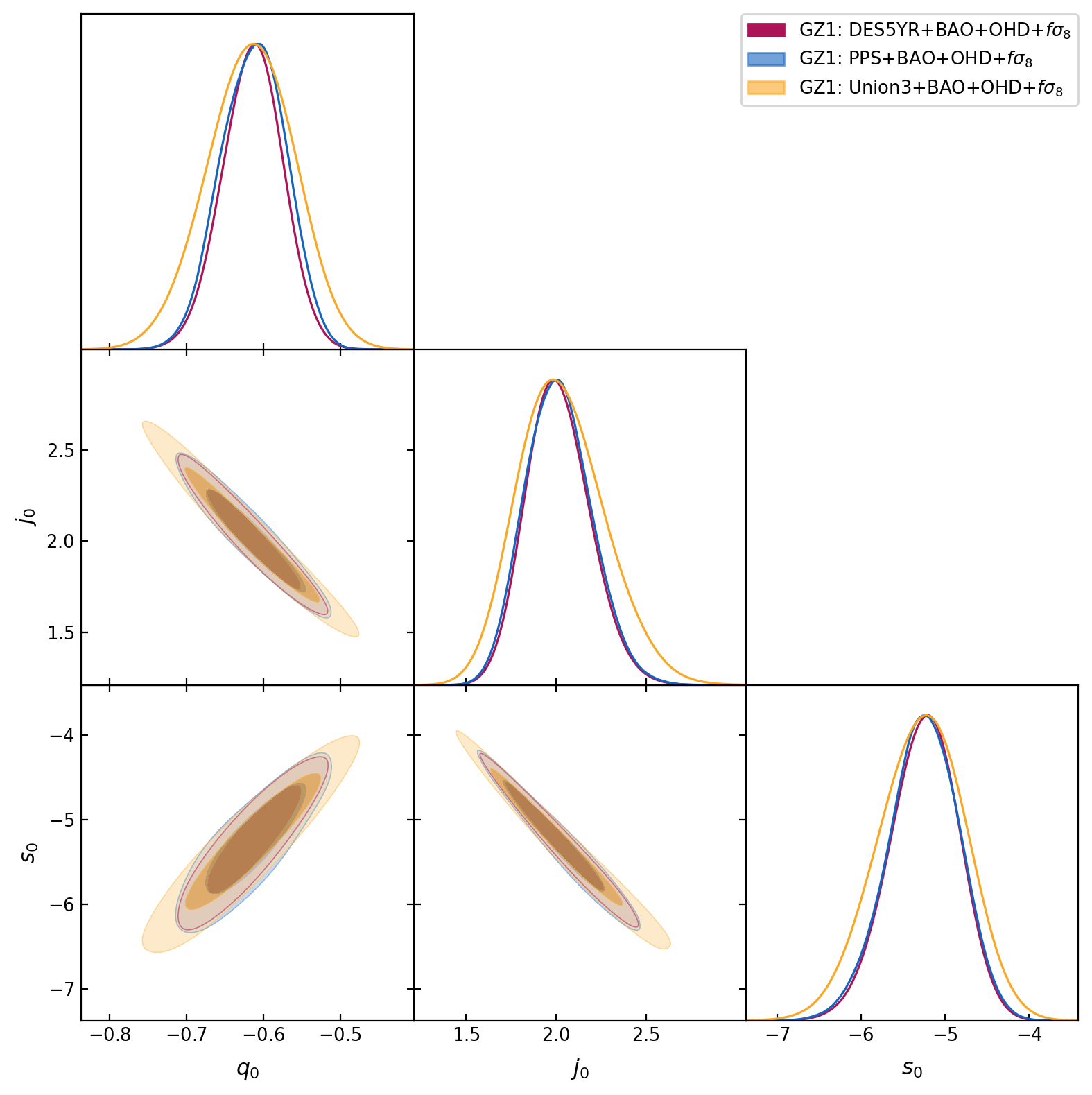}
%        \caption*{(a) Cosmographic analysis for the GZ1 model}
    \end{minipage}
    \hfill
    \begin{minipage}[b]{0.48\textwidth}
        \centering
        \includegraphics[width=\textwidth]{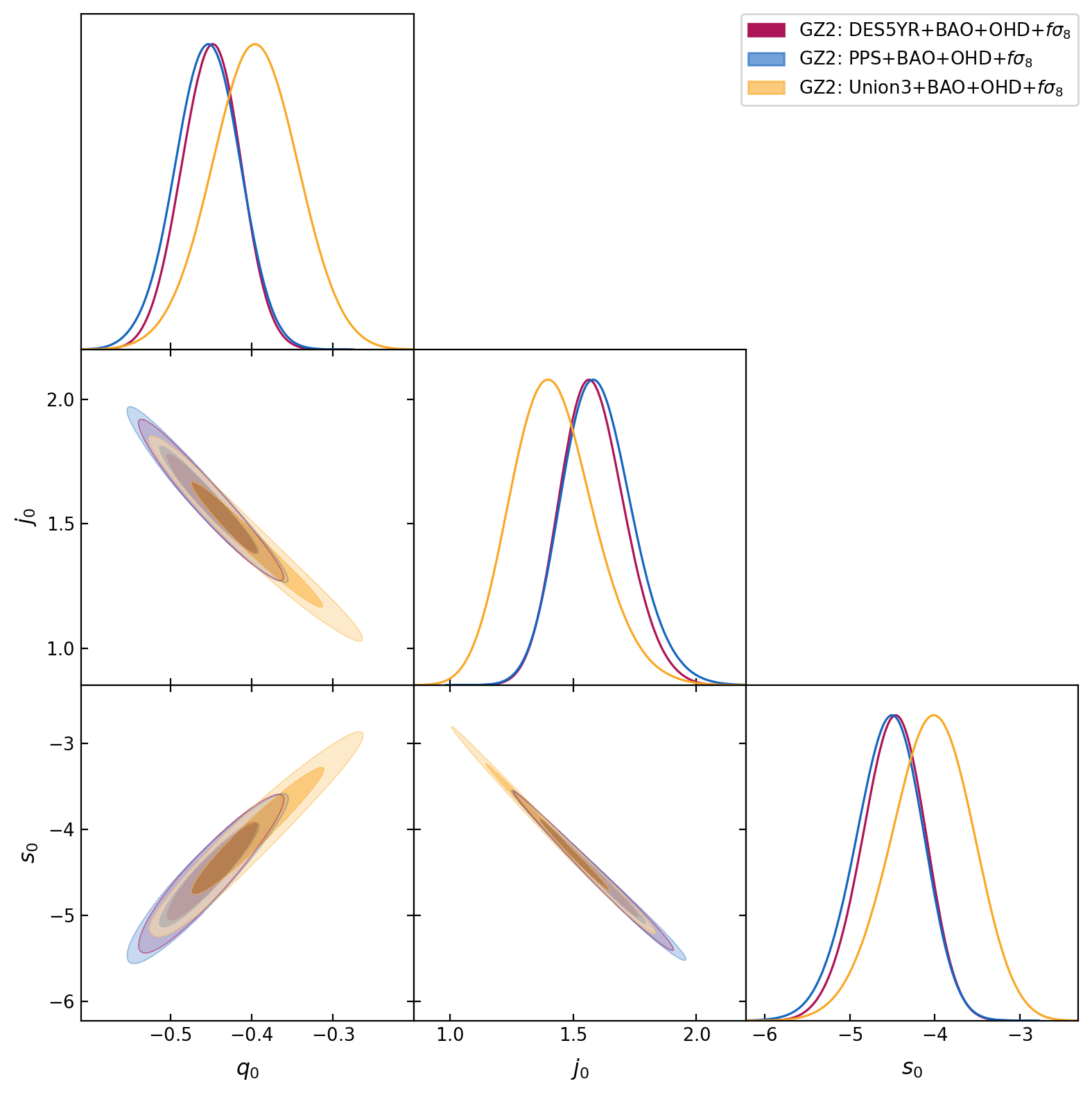}
%        \caption*{(b) Cosmographic analysis for the GZ2 model}
    \end{minipage}

    \caption{Cosmographic corner plots showing the marginalized posterior distributions and parameter correlations for the GZ1 (left) and GZ2 (right) models.}
    \label{fig:cg-GZ}
\end{figure}

The numerical values of the cosmographic parameters inferred from the MCMC
analysis are summarized in Table~\ref{tab:tab2} for the different dataset
combinations. Figure~\ref{fig:cg-GZ} displays the corresponding cosmographic corner plots for the GZ1 and GZ2 models, illustrating the marginalized posterior distributions and joint constraints in the $(q_0,j_0,s_0)$ parameter space. In both parametrizations, the deceleration parameter $q_0$ is tightly constrained and consistently remains negative, providing strong evidence for the present accelerated expansion of the Universe. However, noticeable differences appear in the higher--order cosmographic parameters. The GZ1 model prefers relatively larger values of $j_0$ together with more negative values of $s_0$, accompanied by strong correlations among the parameter pairs $(q_0,j_0)$, $(q_0,s_0)$, and $(j_0,s_0)$. This behaviour points toward a comparatively richer and more dynamically evolving late--time expansion history. In contrast, the GZ2 parametrization exhibits posterior distributions shifted toward smaller $j_0$ and comparatively less negative $s_0$, with smoother and tighter parameter degeneracies, indicating a milder deviation from the standard $\Lambda$CDM cosmographic behaviour. The trends observed in the corner plots are fully consistent with the numerical results reported in Table~\ref{tab:tab2}\\.

\begin{figure}[h!]
    \centering
        \includegraphics[width=0.8\textwidth]{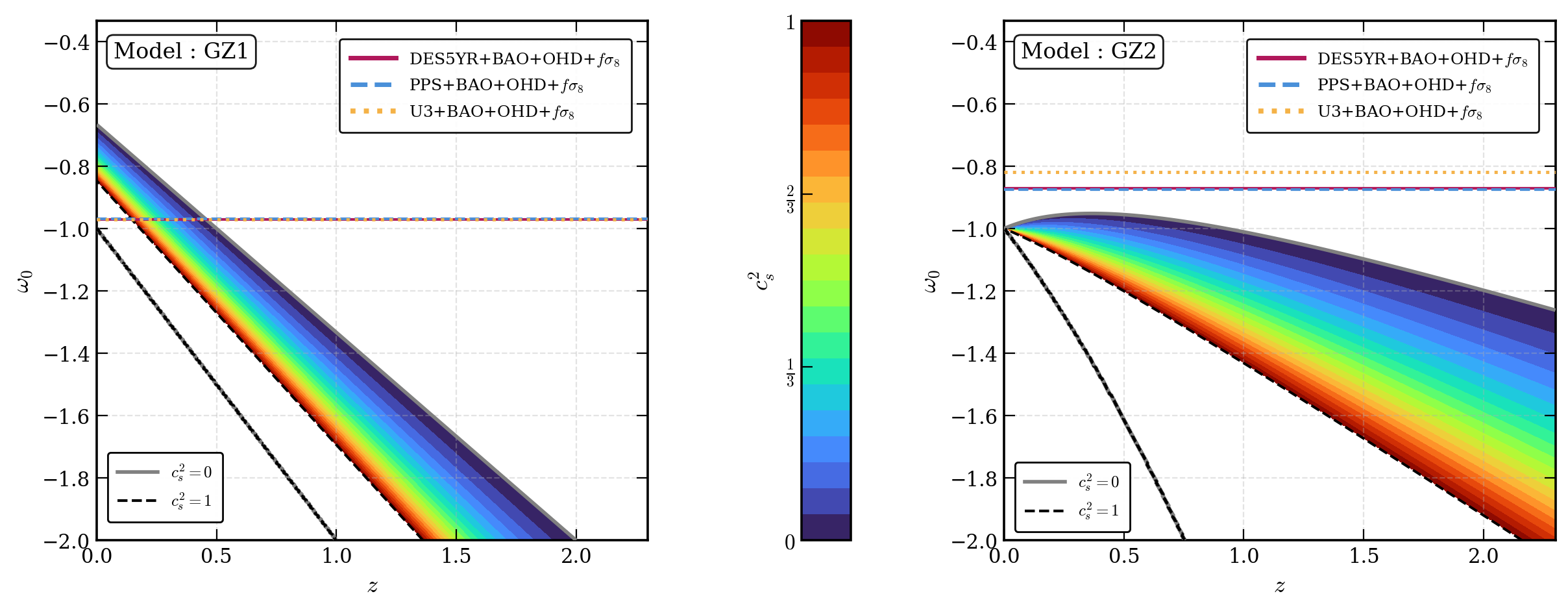}
            \caption{ 
Adiabatic sound speed squared $c_s^2$ for the Gong--Zhang Type~I (left) and Type~II
(right) models.
Shaded regions denote the stable and causal domain $0<c_s^2<1$ in $(z,\omega_0)$
plane, bounded by the curves $c_s^2=0$ and $c_s^2=1$.}
    \label{fig:cs2_gz}
\end{figure}

Figure~\ref{fig:cs2_gz} illustrates the redshift and parameter dependence of the adiabatic sound speed squared, $c_s^2=\partial p/\partial\rho$, for the Gong--Zhang Type~I (GZ1, left) and Type~II (GZ2, right) dark--energy parametrizations in the $(z,\omega_0)$ plane.
For the GZ1 model, 
\begin{align}
c_s^2 &=\frac{2\omega_0(1+z)+3\omega_0^2}{3(1+z)^2+3\omega_0(1+z)}
\end{align}
  the region satisfying the physical conditions $0<c_s^2<1$ is confined to a relatively narrow wedge that progressively contracts toward higher redshift and more negative $\omega_0$, indicating a rapid loss of perturbative stability away from the late--time regime. In contrast,   the GZ2 parametrization,
  \begin{align}
c_s^2 &=\frac{2\omega_0(1+z)e^{z/(1+z)}
+\omega_0\!\left(1+3\omega_0 e^{z/(1+z)}\right)e^{z/(1+z)}}
{3(1+z)^2+3\omega_0 e^{z/(1+z)}}
\end{align}, 
admits a broader and more gradually evolving domain with positive sound speed, reflecting the smoother redshift dependence induced by the exponential modulation.
The curves $c_s^2=0$ and $c_s^2=1$ delineate the boundaries between unstable, causal, and superluminal regimes, while the horizontal lines denote the observationally preferred best--fit values of $\omega_0$ inferred from background expansion data. Taken together, these features demonstrate that perturbative stability provides a complementary consistency criterion, imposing stronger restrictions on the
GZ1 parameter space and comparatively milder constraints on GZ2, particularly at moderate redshifts.\\

\begin{figure}[h!]
    \centering
    \includegraphics[width=0.8\textwidth]{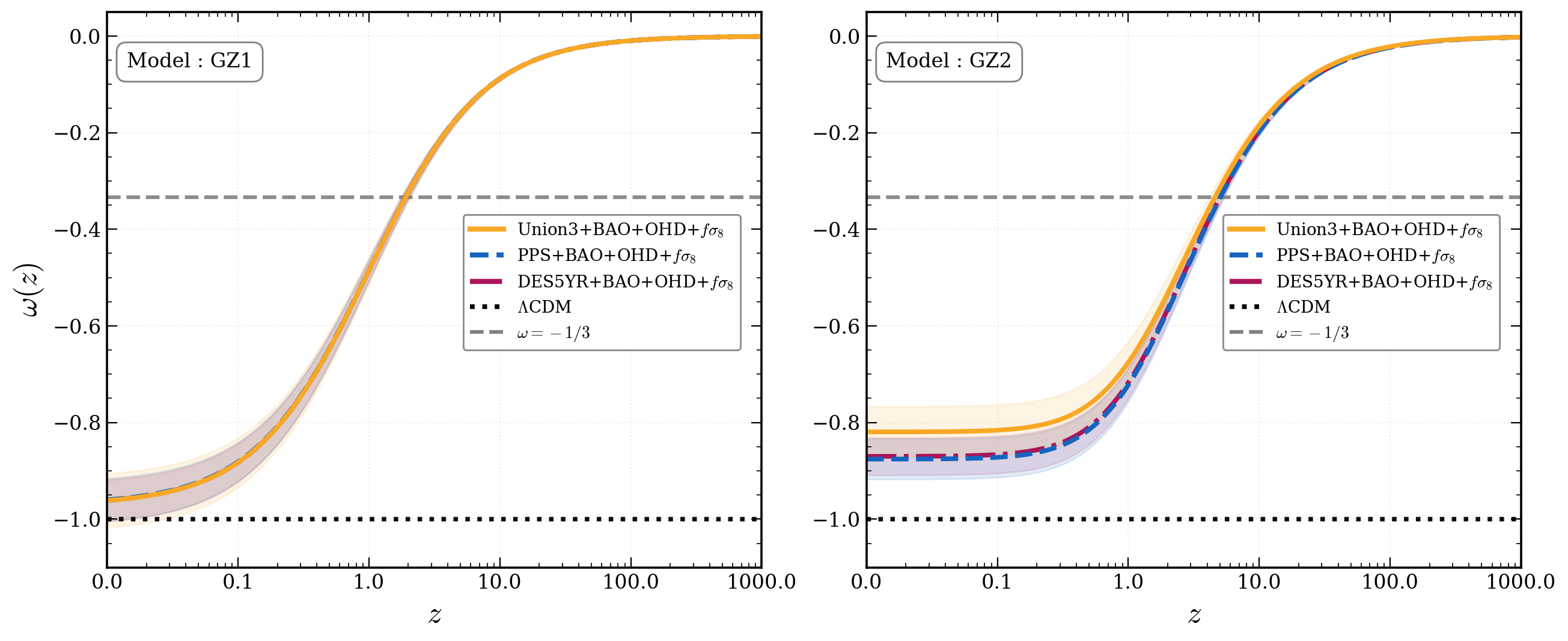}

    \caption{
    Evolution of the dark energy equation of state parameter $\omega(z)$  for the GZ1 (left panels) and GZ2 (right panels) dark energy parametrizations as functions of redshift. The reconstructed evolutions are obtained using the best-fit parameters from the combined late-time datasets and shaded regions represent the corresponding 1-$\sigma$ uncertainties.
    }

    \label{fig:eos}
\end{figure}

Figure~\ref{fig:eos} shows the evolution of the dark energy equation of state parameter $\omega(z)$ for the GZ1 and GZ2 parametrizations over a wide redshift range up to $z=1000$. In both models, the equation of state gradually evolves toward $\omega(z)\rightarrow 0$ at high redshift, indicating that the dark energy component acquires a matter-like behaviour in the early Universe. This feature is a direct consequence of the functional form of the Gong--Zhang parametrizations and reflects their phenomenological nature. However, it is important to emphasize that these models are primarily constructed to describe the late-time dynamics of cosmic acceleration rather than the complete evolution of the Universe across all epochs. In the observationally relevant low- and intermediate-redshift regime, both GZ1 and GZ2 remain smooth and well behaved, while the inclusion of the $f\sigma_8$ growth data already restricts large deviations from the standard cosmological evolution. The figure also shows that the transition from the accelerating regime ($\omega<-1/3$) to the matter-like regime occurs only gradually at sufficiently high redshift. Therefore, the asymptotic behaviour $\omega(z)\to0$ should be interpreted as a limitation of the parametrization at very early times rather than as a direct inconsistency of the models for late-time cosmology, where they continue to provide viable effective descriptions of dark energy dynamics. So the present analysis should be interpreted primarily as a late–time effective description constrained by low–redshift observations, rather than a complete physical model applicable to the recombination epoch or primordial cosmology.

\begin{figure}[h!]
    \centering
    \includegraphics[width=0.8\textwidth]{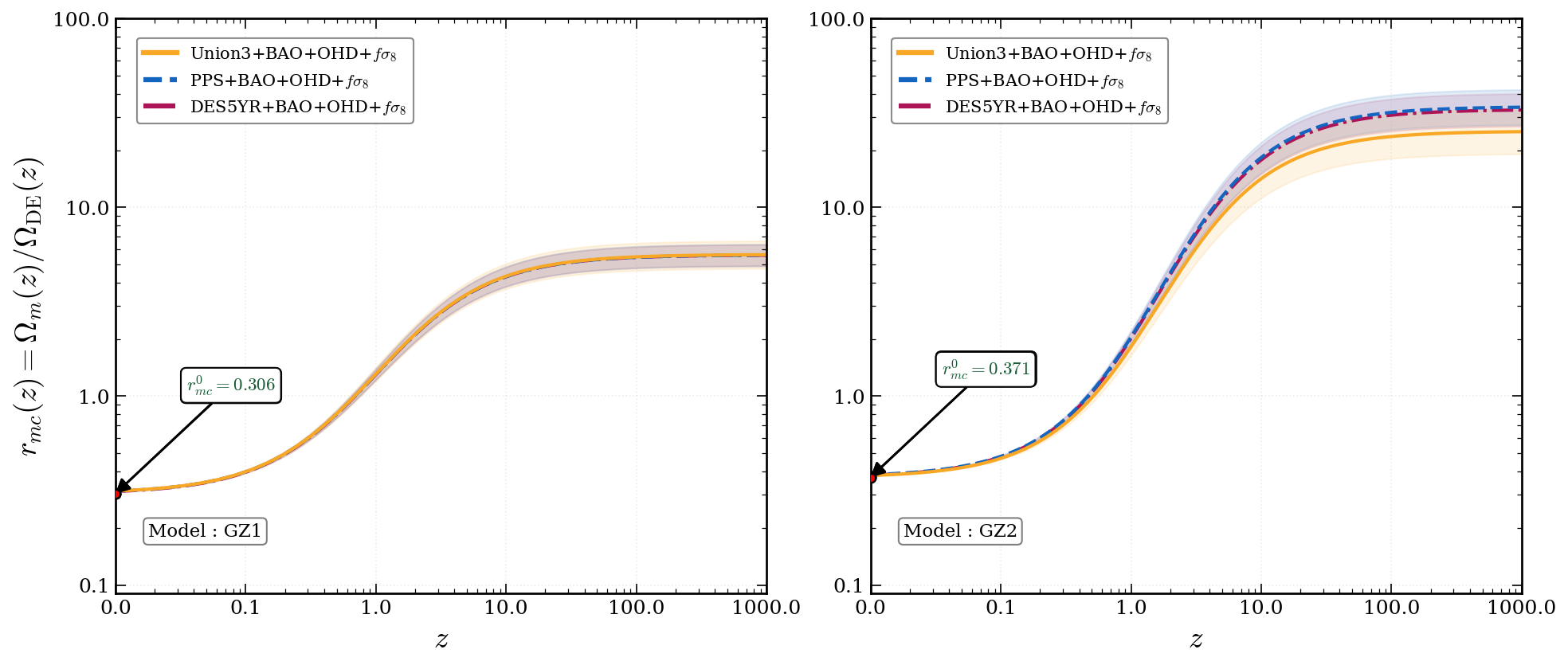}
    \caption{Evolution of the coincidence parameter for the GZ1 (left) and GZ2 (right) parametrizations using the best-fit parameters from the Union3+BAO+OHD+$f\sigma_8$, PPS+BAO+OHD+$f\sigma_8$, and DES5YR+BAO+OHD+$f\sigma_8$ datasets. The shaded regions represent the corresponding 1-$\sigma$ uncertainties, while the annotated markers indicate the present-epoch values.}

    \label{fig:rmc}
\end{figure}

The evolution of the coincidence parameter,
$r_{mc}(z)\equiv \frac{\Omega_m(z)}{\Omega_{\rm DE}(z)}$,
where for a spatially flat Universe
$\Omega_{\rm DE}(z)=1-\Omega_m(z)-\Omega_r(z)$,
is shown in Fig.~\ref{fig:rmc} for the GZ1 and GZ2 parametrizations using the best--fit values together with the corresponding $1\sigma$ uncertainties obtained from the Union3+BAO+OHD+$f\sigma_8$, PPS+BAO+OHD+$f\sigma_8$, and DES5YR+BAO+OHD+$f\sigma_8$ dataset combinations. In both models,
$r_{mc}(z)$ remains small near the present epoch, reflecting the
dominance of dark energy at late times, while it gradually increases
toward higher redshifts as the matter component becomes progressively
more dominant. The overall behaviour indicates a smooth recovery of the
standard matter--dominated regime at early times with a suppressed dark
energy contribution. The GZ1 parametrization exhibits an evolution very
close to a $\Lambda$CDM--like behaviour, whereas the GZ2 model shows a
comparatively stronger rise of the coincidence parameter at intermediate
and high redshifts, although still remaining well behaved within the
observationally allowed region. The shaded bands further demonstrate
that the evolution remains tightly constrained across all dataset
combinations, with only mild deviations among the different observational
samples.

\section{Configuration Entropy as a Physical Probe of Structure Growth}
\label{sec:CE}

The growth of cosmic structure is an inherently irreversible process driven by
gravitational instability, during which the matter distribution progressively
departs from large--scale homogeneity.
While the adiabatic sound speed $c_s^2$ characterizes the microphysical
stability of dark--energy perturbations and regulates the propagation of
pressure fluctuations, its imprint on structure formation is ultimately
realized through the clustering of matter.
Configuration entropy provides an information--theoretic and thermodynamic
framework to quantify this macroscopic outcome by tracking how gravitational
instability redistributes matter and accumulates information over cosmic time.
Unlike conventional growth observables that probe only instantaneous amplitudes
or local rates, configuration entropy encodes the integrated history of
structure formation and is therefore sensitive to both the background expansion
and the perturbative properties of the cosmic fluid.
 \subsection{Physical meaning of configuration entropy}
\label{subsec:CE1}
For a pressureless matter fluid with density $\rho(\mathbf{x},t)$ defined over
a comoving volume $V$, the configuration entropy is introduced as a
measure of the spatial information content of the matter distribution.
To ensure a well--defined entropy functional, we work with the normalized
density contrast
$p(\mathbf{x},t)=\rho(\mathbf{x},t)/M$, where
$M=\int_V \rho(\mathbf{x},t)\,dV$ is the conserved total mass.
The configuration entropy is then defined as \cite{Bekenstein:2008smd,Pandey:2017tgy,Das:2018shd,Shannon:1948}
\begin{equation}
S_c(t)
= -\int_V p(\mathbf{x},t)\,\ln p(\mathbf{x},t)\, dV ,
\label{eq:Sc_phys}
\end{equation}
 Up to an additive constant set by the total mass, this definition is equivalent to the Shannon entropy of the normalized density field. This quantity provides a global measure of the degree of spatial disorder in
the matter distribution.
In a perfectly homogeneous universe, the density is uniform,
$p(\mathbf{x},t)=\text{const.}$, and the configuration entropy attains its
maximum value, remaining constant in time.
As gravitational instability amplifies density perturbations and matter
progressively clusters, the effective configuration space available to the
system is reduced, leading to a monotonic decrease of $S_c$.
From a physical perspective, configuration entropy quantifies how efficiently
gravity transforms an initially smooth matter distribution into structured
patterns such as filaments, walls, and halos. Because it encodes the cumulative outcome of nonlinear structure formation,
configuration entropy offers a complementary, thermodynamic description of
cosmic inhomogeneity that goes beyond local or perturbation--based diagnostics.
Unlike standard measures that track the growth of individual modes,
$S_c$ captures the global redistribution of matter and provides a unified
characterization of large--scale structure formation in an expanding universe.

\subsection{Entropy evolution and the role of structure growth}
\label{subsec:CE2}
The evolution of configuration entropy follows directly from mass conservation
and the continuity equation governing a pressureless matter fluid in an
expanding universe.
Expressed in terms of the scale factor $a$ ($a=(1+z)^{-1}$,
where $z$ denotes the cosmological redshift), the entropy evolution equation
takes the form, the entropy evolution can be written
as \cite{Das:2018shd},
\begin{equation}
\frac{dS_c}{da}
= -\frac{3}{a}\bigl(S_c - M\bigr)
- D(a)\,D'(a),
\label{eq:Sc_core}
\end{equation}
where $M=\int_V \rho(\mathbf{x},t)\,dV$ denotes the conserved total mass within
the comoving volume $V$, and $D(a)$ is the linear growth factor of matter density
perturbations, with a prime indicating differentiation with respect to $a$.
When the normalized density $p=\rho/M$ is adopted in the definition of
$S_c$, the quantity $M$ simply sets the constant reference level of entropy and
does not affect the dynamical evolution. Each term in Eq.~\eqref{eq:Sc_core} admits a clear physical interpretation.
The first term represents the dilution effect associated with the cosmological
expansion, reflecting the tendency of the background expansion to preserve
homogeneity.
In contrast, the second term acts as a genuine source term that captures entropy
production driven by the growth of density perturbations.
This contribution encodes the irreversible redistribution of matter induced by
gravitational clustering and establishes a direct connection between structure
formation and the thermodynamic evolution of the cosmic matter distribution. Eq.~\eqref{eq:Sc_core} therefore highlights a central physical result:
the evolution of configuration entropy is directly sourced by the growth of
matter perturbations.
Any physical mechanism that alters the growth history—such as a dynamical
dark--energy equation of state or modified expansion rate—will necessarily leave
a characteristic imprint on the entropy evolution.
Configuration entropy thus provides a global, physically motivated diagnostic
that links late--time structure formation to the underlying cosmological
dynamics.

\subsection{Numerical implementation and entropy diagnostics}
\label{subsec:NME}
Within the Gong--Zhang framework, dynamical dark energy influences configuration entropy through the coupled evolution of the background expansion rate $E(a)=H(a)/H_0$ and the linear growth factor $D(a)$. The GZ1 and GZ2 parametrizations, characterized by $\omega(a)=\omega_0a$ and $\omega(a)=\omega_0ae^{1-a}$ respectively, modify the late--time expansion and growth histories while keeping dark energy dynamically suppressed at early times ($a\ll1$), thereby preserving an effective matter--dominated behaviour. The growth evolution, quantified through the logarithmic growth rate $f(a)=d\ln D/d\ln a$, directly determines the entropy production history and allows deviations from $\Lambda$CDM to be characterized through the relative growth--rate deviation $\Delta f(a)$, which is already discussed in \ref{subsec:GZ_background_growth}.\\

The entropy evolution is evaluated by fixing the conserved mass $M$ and imposing
initial conditions deep in the matter--dominated era ($a_i\ll1$), where dark--energy
effects are negligible.
We normalize the configuration entropy to $S_c(a_i)=1$ and present all results in
terms of the ratios $S_c(a)/S_c(a_i)$ or $\Delta R(a)$, thereby removing any
dependence on the arbitrary entropy normalization.
The linear growth factor is initialized with $D(a_i)\propto a_i$ and normalized to
unity at the present epoch, $D(1)=1$, ensuring standard early--time growth.
With these choices, the entropy diagnostics isolate genuine late--time physical
effects induced by dynamical dark energy.\\

The coupled evolution of the linear growth factor and the configuration entropy
is solved numerically, with initial conditions imposed deep in the
matter--dominated era to ensure insensitivity to early--time transients.
The growth factor is normalized to unity at the present epoch, $D(a=1)=1$,
and we track the normalized entropy ratio $S_c(a)/S_c(a_i)$ in order to remove
any dependence on the arbitrary normalization of the initial entropy.  A particularly informative diagnostic of the thermodynamic impact of structure
formation is the entropy--production rate,
\begin{equation}
R(a) \equiv \frac{d\ln S_c}{da},
\end{equation}
which quantifies the instantaneous rate at which information is redistributed
by gravitational clustering during cosmic evolution.
To isolate the effect of dynamical dark energy, we further define the relative
entropy--rate deviation with respect to the standard cosmological model,
\begin{equation}
\Delta R(a) \equiv
R_{\rm model}(a) - R_{\Lambda{\rm CDM}}(a),
\end{equation}
thereby providing a direct, purely thermodynamic comparison between competing
cosmological scenarios.
This quantity highlights deviations from $\Lambda$CDM in the late--time regime
and serves as a sensitive probe of how modifications to the expansion history
and growth of structure imprint themselves on the global entropy evolution of
the Universe.

\subsection{Results and physical interpretation}
\label{subsec:CE_results}
In this subsection, all relevant numerical results are generated using the best--fit parameter values obtained from the MCMC analysis listed in Table~\ref{tab:tab1}.\\

\begin{figure}[h!]
    \centering
    
    \begin{minipage}[b]{0.48\textwidth}
        \centering
        \includegraphics[width=\textwidth]{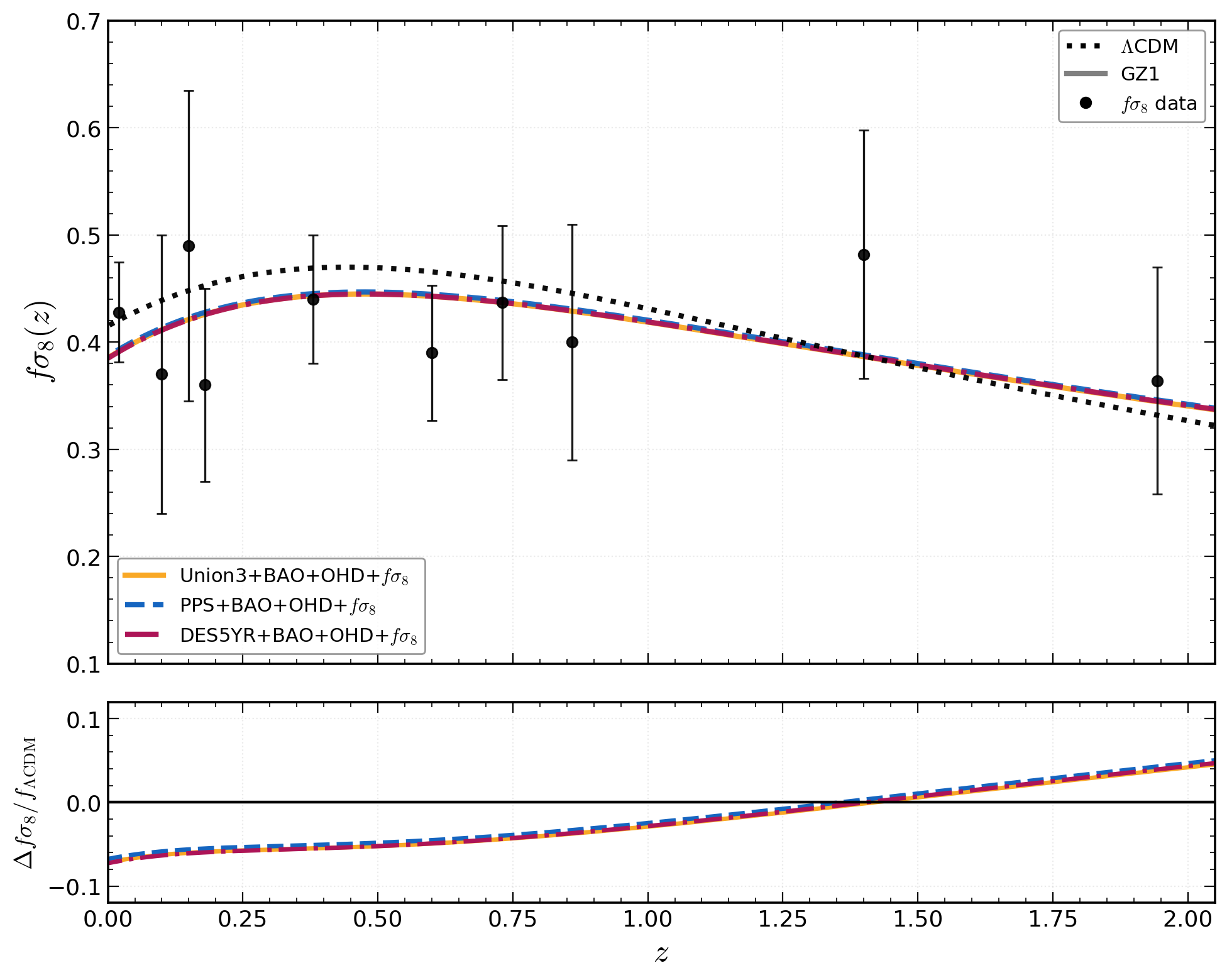}
    \end{minipage}
    \hfill
    \begin{minipage}[b]{0.48\textwidth}
        \centering
        \includegraphics[width=\textwidth]{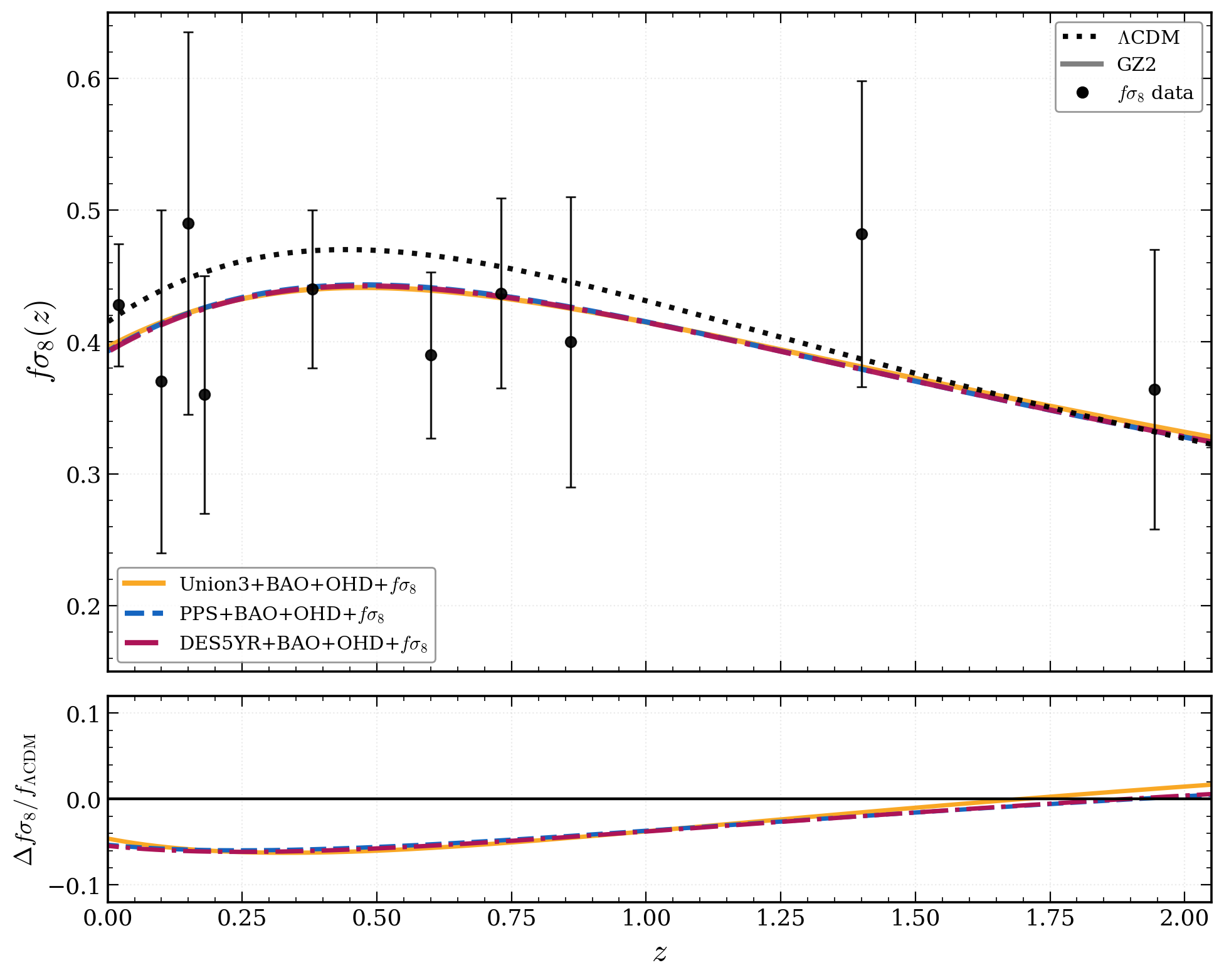}
    \end{minipage}

    \caption{
    Evolution of the growth-rate observable $f\sigma_8(z)$ for the GZ1 (left panel) and GZ2 (right panel) dark energy parametrizations in comparison with the $\Lambda$CDM model as a function of redshift for different SN$+$BAO$+$OHD$+f\sigma_8$ dataset combinations. The black points with error bars represent the observational $f\sigma_8$ measurements, while the lower subpanels show the relative deviation, $\Delta f\sigma_8/f\sigma_{8,\Lambda{\rm CDM}}$, with respect to the standard $\Lambda$CDM scenario.
    }
    
    \label{fig:fs8}
\end{figure}

Figure~\ref{fig:fs8} shows that both the GZ1 and GZ2 models reproduce the observed evolution of the growth-rate parameter $f\sigma_8(z)$ remarkably well and remain consistent with the current large-scale structure observations, including the recently released DESI-DR2 growth measurements with their corresponding error bars. The observational points are presented as functions of redshift, since the availability and uncertainties of the measured $f\sigma_8$ data are conventionally expressed in terms of $z$. This demonstrates that the process of structure and galaxy cluster formation in these models proceeds in a manner very similar to the standard $\Lambda$CDM cosmology. At the same time, small deviations from $\Lambda$CDM appear at late times, where both parametrizations predict a slightly lower growth amplitude. The lower panels further show that these deviations remain only at the level of a few percent throughout the observed redshift range. Among the two models, GZ2 exhibits a comparatively stronger suppression of growth than GZ1, indicating a mildly reduced clustering tendency and a slower late-time growth of matter perturbations relative to the $\Lambda$CDM model. For the remaining analyses in this section, the cosmological evolution is presented in terms of the scale factor $a$, using the relation $1/a = 1+z$, which provides a more convenient description of the dynamical evolution across cosmic time.\\

\begin{figure}[h!]
    \centering
    \begin{minipage}[b]{0.48\textwidth}
        \centering
        \includegraphics[width=\textwidth]{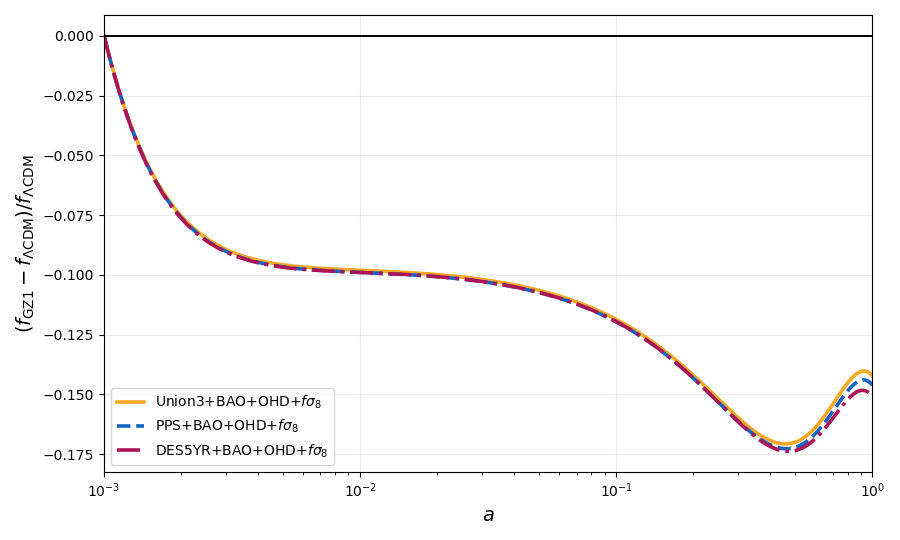}
%        \caption*{(a) $\Delta R_{\rm LCDM-GZ1}$ at $M=2$, $S_c(a_i)=1$}
    \end{minipage}
    \hfill
    \begin{minipage}[b]{0.48\textwidth}
        \centering
        \includegraphics[width=\textwidth]{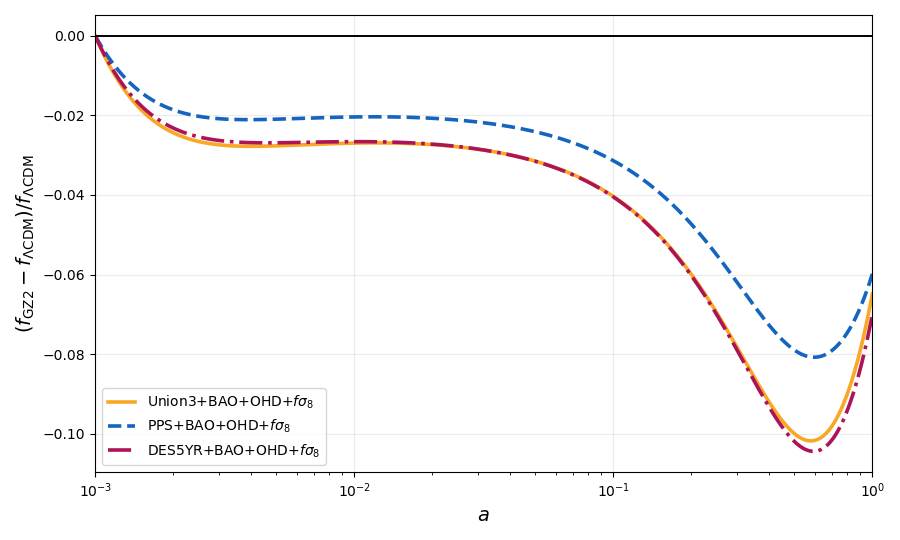}
%        \caption*{(b) $\Delta R_{\rm LCDM-GZ2}$ at $M=2$, $S_c(a_i)=1$}
    \end{minipage}

 \caption{Relative growth--rate deviation
$\left[f_{\rm GZ}-f_{\Lambda{\rm CDM}}\right]/f_{\Lambda{\rm CDM}}$
for the Gong--Zhang models using SN+BAO+OHD+f$\sigma_8$ data.
The left panel corresponds to the Type~I (GZ1) parametrization,
while the right panel shows the Type~II (GZ2) case,
illustrating suppressed late--time structure growth with standard
early--time behavior.}
 
    \label{fig:f}
\end{figure}

From Fig.~\ref{fig:f}, the relative growth--rate deviation $\Delta f(a)$ remains very close to zero at early times ($a\ll1$), indicating that both Gong--Zhang parametrizations successfully recover the standard $\Lambda$CDM growth behaviour during the radiation-- and matter--dominated epochs. As the Universe evolves and dark energy becomes dynamically important, $\Delta f(a)$ gradually departs from zero and acquires negative values, reflecting a suppression of structure growth relative to $\Lambda$CDM due to the evolving dark--energy equation of state. The suppression becomes more pronounced at intermediate and late times, particularly for the GZ2 parametrization, which exhibits comparatively larger deviations from the standard growth history. Near the present epoch, the curves show a mild upward recovery, indicating that residual matter clustering still partially counteracts the accelerated expansion driven by dark energy. Overall, the smooth evolution of $\Delta f(a)$ across all dataset combinations demonstrates that the Gong--Zhang parametrizations modify the late--time growth history in a controlled and physically consistent manner without introducing instabilities or unphysical behaviour.\\

\begin{figure}[h!]
    \centering
    \begin{minipage}[b]{0.32\textwidth}
        \centering
        \includegraphics[width=\textwidth]{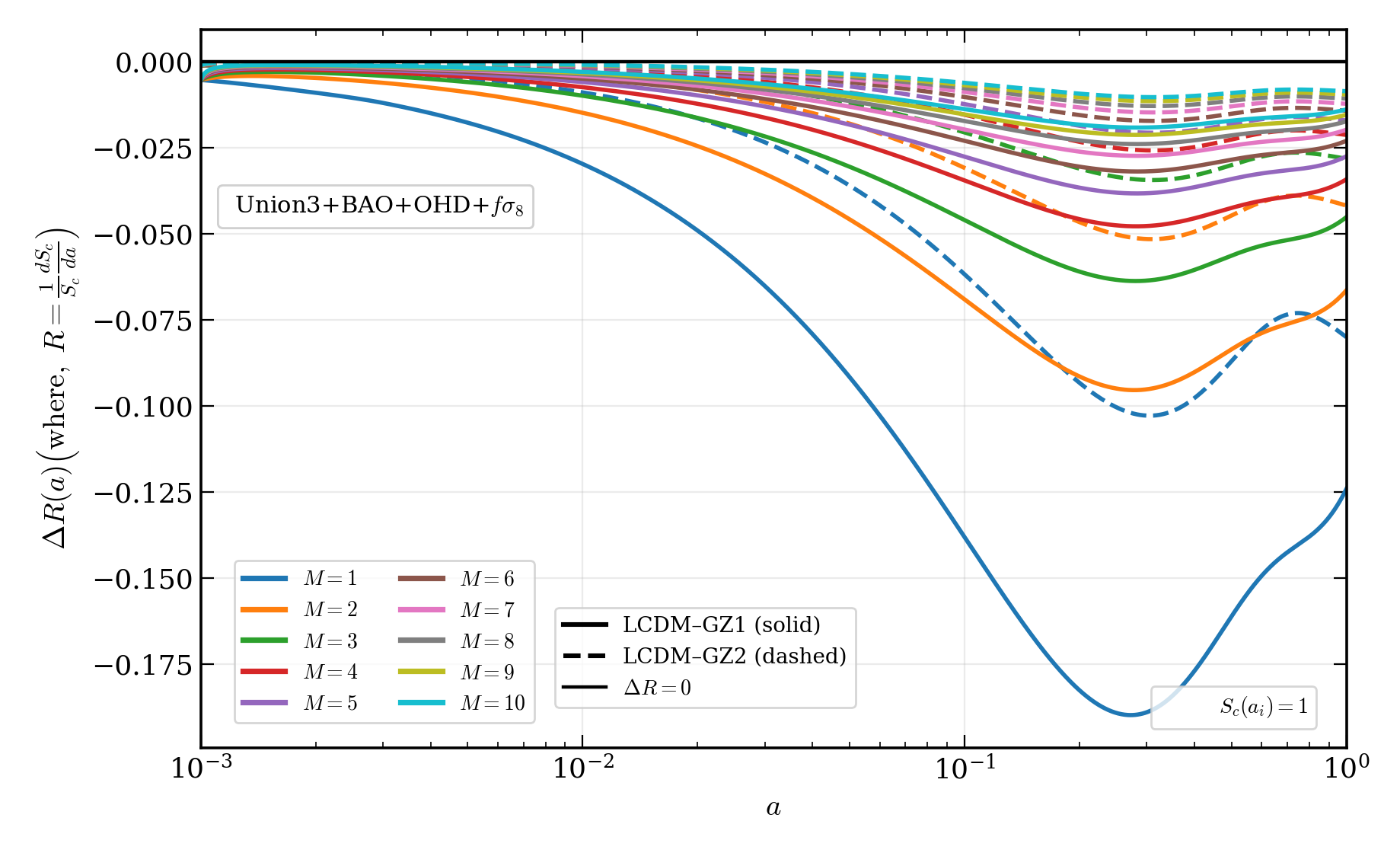}
%        \caption*{(a) $\Delta R(a)$ for \textbf{Union3} + BAO + OHD}
    \end{minipage}
    \hfill
    \begin{minipage}[b]{0.32\textwidth}
        \centering
        \includegraphics[width=\textwidth]{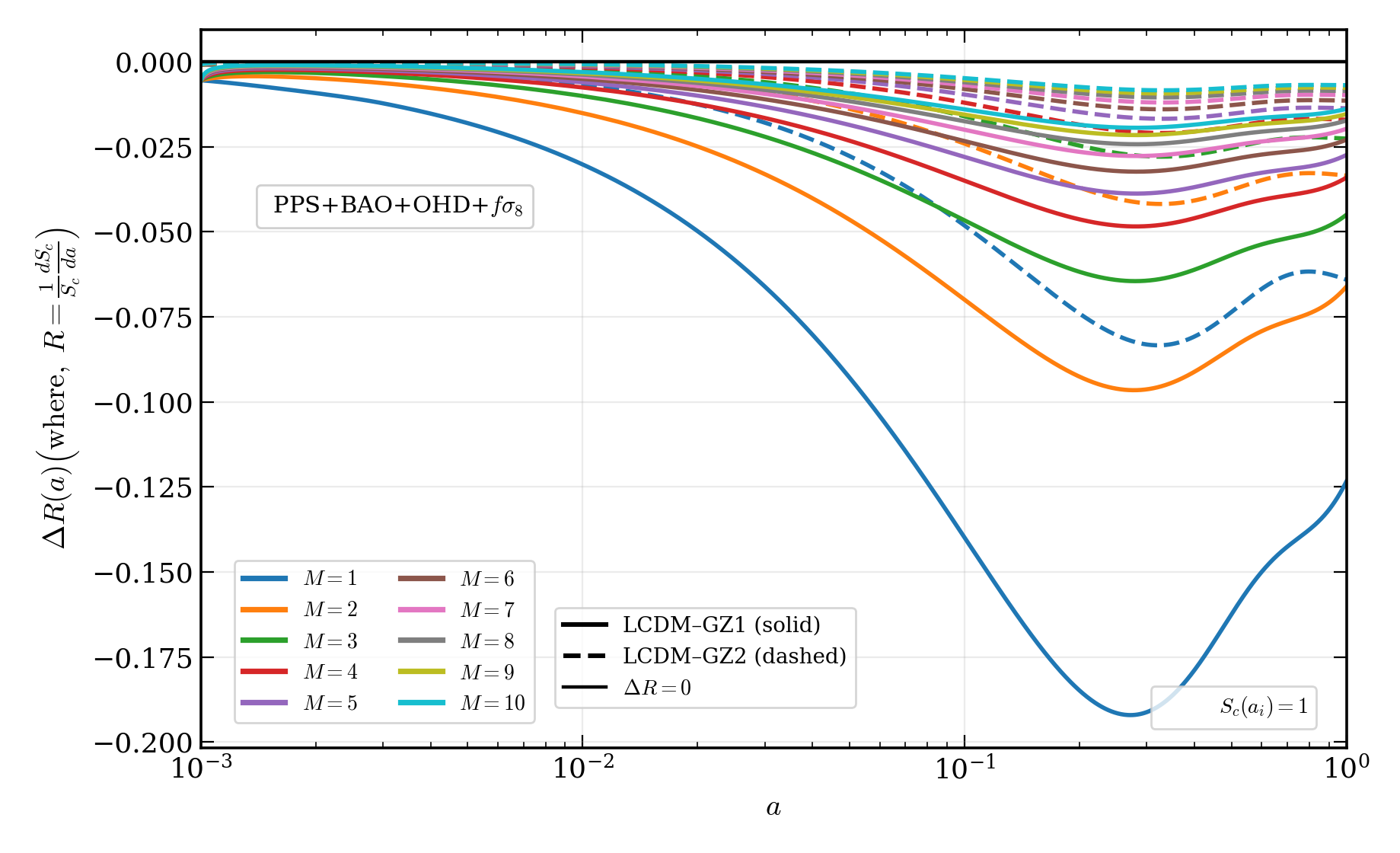}
%        \caption*{(b) $\Delta R(a)$ for \textbf{Pantheon+SH0ES} + BAO + OHD}
    \end{minipage}
    \hfill
    \begin{minipage}[b]{0.32\textwidth}
        \centering
        \includegraphics[width=\textwidth]{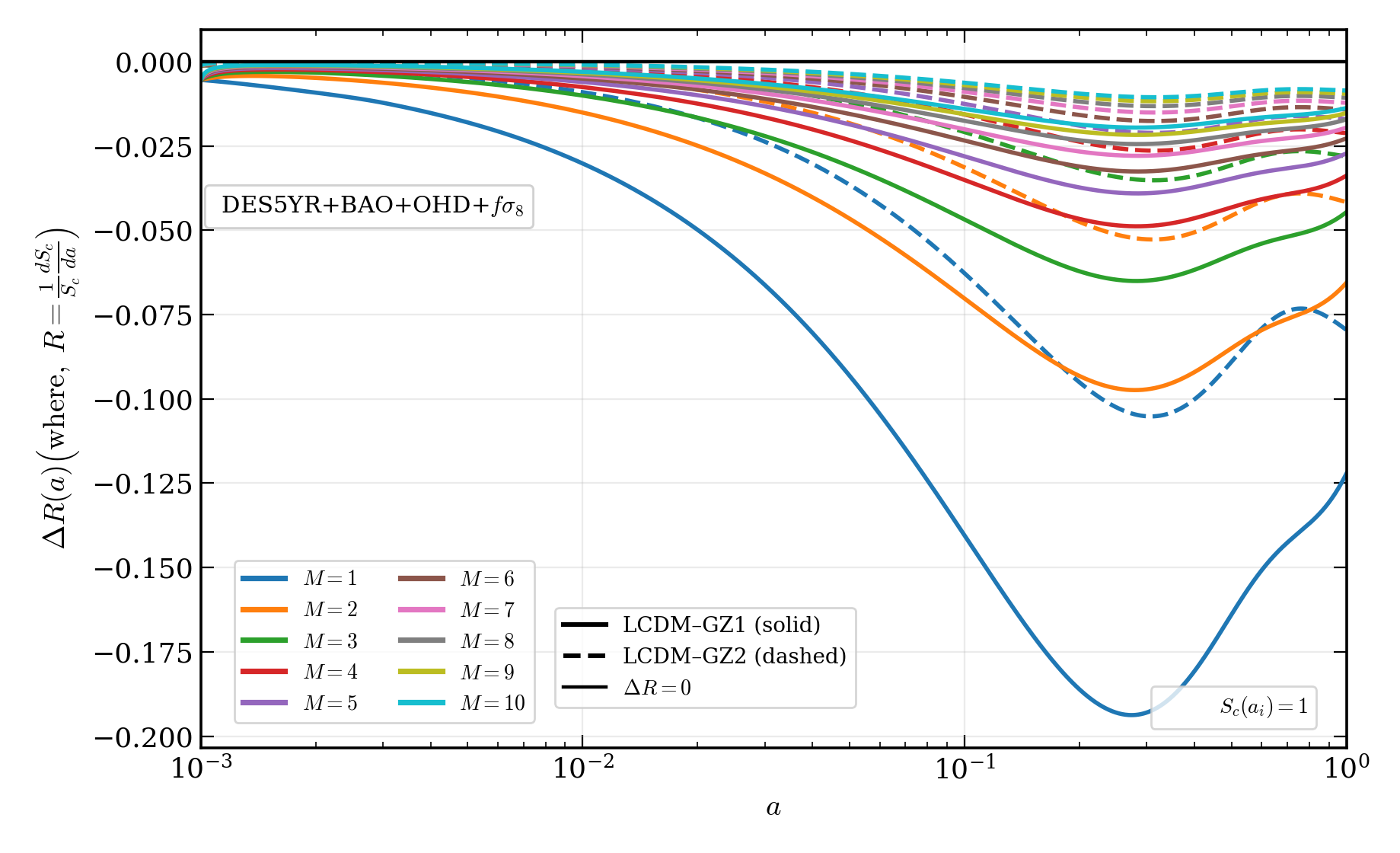}
%%        \caption*{(c) $\Delta R(a)$ for \textbf{DES--SN5YR} + BAO + OHD}
    \end{minipage}

     \caption{Evolution of the entropy–production–rate deviation $\Delta R(a)$ for the Gong--Zhang models across different SN+BAO+OHD+$f\sigma_8$ dataset combinations: (a) Union3+BAO+OHD+$f\sigma_8$, (b) Pantheon+SH0ES+BAO+OHD+$f\sigma_8$, and (c) DES--SN5YR+BAO+OHD+$f\sigma_8$.}
 
    \label{fig:CE1}
\end{figure}

\begin{figure}[h!]
    \centering
    \begin{minipage}[b]{0.48\textwidth}
        \centering
        \includegraphics[width=\textwidth]{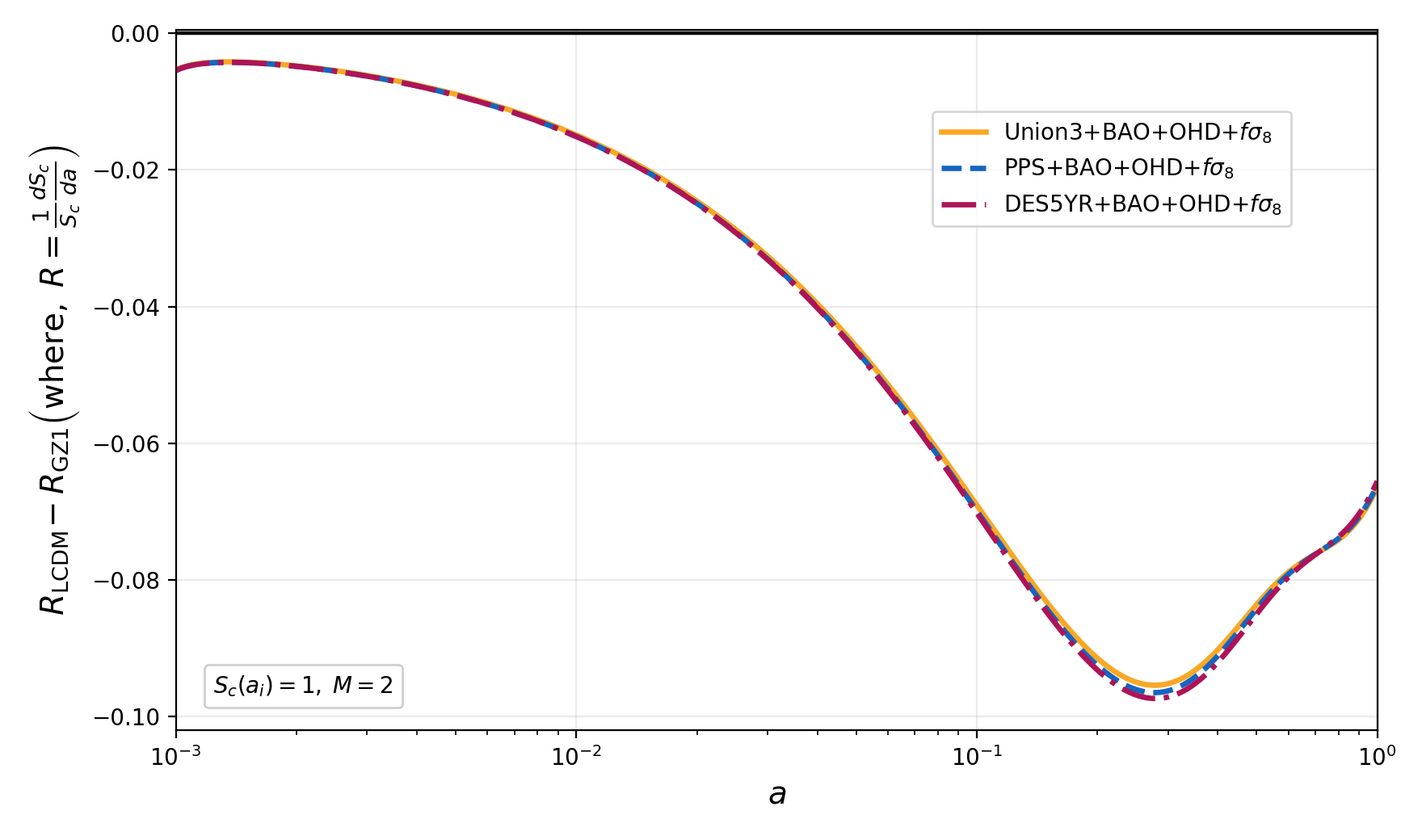}
%        \caption*{(a) $\Delta R_{\rm LCDM-GZ1}$ at $M=2$, $S_c(a_i)=1$}
    \end{minipage}
    \hfill
    \begin{minipage}[b]{0.48\textwidth}
        \centering
        \includegraphics[width=\textwidth]{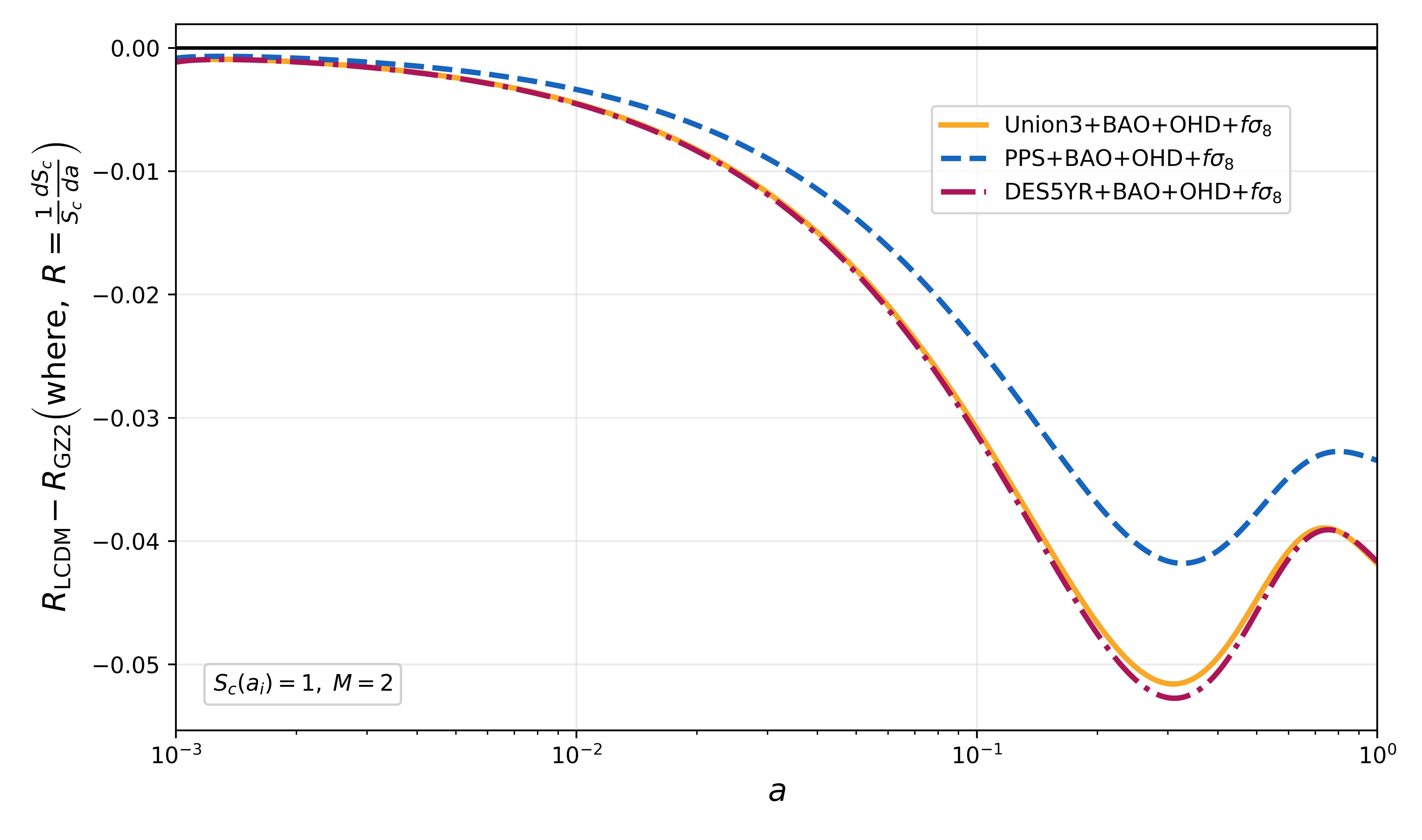}
%        \caption*{(b) $\Delta R_{\rm LCDM-GZ2}$ at $M=2$, $S_c(a_i)=1$}
    \end{minipage}

 \caption{Evolution of  entropy–production–rate deviation  $\Delta R(a)$ relative
to $\Lambda$CDM for the GZ1 (left) and GZ2 (right) models as a function of
the scale factor, shown for $M=2$ and $S_c(a_i)=1$ using different
SN+BAO+OHD+$f\sigma_8$ dataset combinations.}
 
    \label{fig:CE2}
\end{figure}

 Figures~\ref{fig:CE1} and \ref{fig:CE2} summarize the main thermodynamic implications of the configuration--entropy analysis for the Gong--Zhang dark--energy parametrizations. The relative entropy--production--rate deviation, $\Delta R(a)\equiv R_{\rm model}(a)-R_{\Lambda{\rm CDM}}(a)$, provides a direct measure of how the entropy generation associated with structure formation differs from the standard $\Lambda$CDM scenario. As shown in Fig.~\ref{fig:CE1}, both Gong--Zhang models exhibit deviations that remain very small at early times, confirming that the dark--energy sector stays dynamically suppressed during the radiation-- and matter--dominated eras and therefore preserves the standard early--time growth history. The near overlap of the curves for different conserved mass scales $M$ further demonstrates that the entropy evolution is largely insensitive to the specific choice of comoving mass in the linear regime. As the Universe evolves toward the late--time accelerated phase, $\Delta R(a)$ gradually departs from zero and becomes negative, indicating a suppression of entropy production relative to $\Lambda$CDM. This behaviour reflects the reduced efficiency of gravitational clustering caused by the evolving dark--energy equation of state. The suppression becomes increasingly significant at intermediate and late times, particularly for smaller conserved mass scales, while still maintaining a smooth and well--behaved evolution across all dataset combinations. Near the present epoch, the curves exhibit a mild recovery tendency, analogous to the behaviour observed in the growth--rate deviation, suggesting that residual matter clustering continues to contribute to entropy production even in the dark--energy dominated regime. Overall, the smooth and synchronized evolution of $\Delta R(a)$ across different values of $M$ demonstrates that the Gong--Zhang parametrizations modify the late--time thermodynamic behaviour of the Universe in a controlled and physically consistent manner without introducing abrupt transitions or instabilities.\\

Figure~\ref{fig:CE2} presents a direct comparison between the entropy--production rates of the Gong--Zhang parametrizations and the standard $\Lambda$CDM scenario through the quantity $\Delta R_{\Lambda{\rm CDM}-{\rm GZ}}(a)$. For both GZ1 and GZ2, the deviation remains very small at early times, indicating that the standard entropy evolution associated with an effective matter--dominated expansion behaviour is approximately reproduced within the late--time phenomenological framework considered here. As the scale factor increases and dark energy begins to dominate the cosmic expansion, the deviation gradually becomes more negative, reflecting a suppression of entropy production relative to $\Lambda$CDM due to the reduced efficiency of gravitational clustering. Both parametrizations exhibit a characteristic late--time minimum followed by a mild recovery toward the present epoch, demonstrating that residual matter clustering continues to contribute to entropy generation even during the accelerated expansion phase.  The GZ2 parametrization exhibits comparatively larger deviations and a shallower recovery relative to GZ1, indicating a stronger modification of the late--time growth and entropy evolution driven by its dynamically evolving equation of state. In contrast, the GZ1 model remains closer to the $\Lambda$CDM behaviour throughout the entire evolution. The close agreement among the different SN+BAO+OHD+$f\sigma_8$ dataset combinations further demonstrates the robustness and stability of the predicted entropy evolution. Overall, the smooth and well--behaved evolution of $\Delta R(a)$ highlights that the Gong--Zhang parametrizations provide controlled and observationally consistent deviations from $\Lambda$CDM while preserving the standard early--time cosmological behaviour.\\

Overall, the evolution of $\Delta R(a)$ highlights configuration entropy as a
physically transparent and thermodynamically motivated probe of dynamical dark
energy, providing a complementary consistency test for the Gong--Zhang models
beyond standard background and perturbative constraints.

\section{Conclusion}
\label{sec:con}

In this work we have performed a comprehensive observational, dynamical, and thermodynamic investigation of the Gong--Zhang Type~I (GZ1) and Type~II (GZ2) dark--energy parametrizations using three independent combinations of late--time cosmological probes, namely Union3+BAO+OHD+$f\sigma_8$, Pantheon+SH0ES+BAO+OHD+$f\sigma_8$, and DES--SN5YR+BAO+OHD+$f\sigma_8$. By marginalizing over the sound--horizon scale and avoiding direct CMB priors, we adopted a fully phenomenological late--time framework designed to isolate the impact of dynamical dark energy on the recent expansion history of the Universe. The resulting marginalized one-- and two--dimensional posterior distributions demonstrate that both parametrizations provide statistically stable and observationally consistent fits to all dataset combinations, yielding robust constraints on the parameter set $\{\Omega_m,\,h,\,h\,r_d,\,\omega_0,\,\sigma_8\}$. In particular, the inferred values of $\Omega_m$ and $h$ remain fully compatible with standard late--time cosmology, while the equation--of--state parameter $\omega_0$ departs from the cosmological--constant limit in a controlled and physically interpretable manner.\\

{\color{darkblue}The correlation matrices reveal the dominant degeneracy directions among the matter sector, distance scale, and dark--energy dynamics. Both parametrizations exhibit the expected strong correlation between $h$ and $h,r_d$, reflecting the manner in which BAO measurements constrain the cosmological distance scale, together with a pronounced anti--correlation between $\Omega_m$ and $\omega_0$. However, these degeneracies are systematically weaker for the GZ2 parametrization, indicating improved separability between the matter and dark--energy sectors and resulting in tighter posterior constraints. This behaviour is reflected in the information--criteria analysis, where the Jeffreys--scale comparison shows that the statistical conclusions depend on both the dataset combination and the adopted information criterion. While both GZ1 and GZ2 generally yield lower AIC values than $\Lambda$CDM, the corresponding BIC values indicate that the preference is less conclusive once model complexity is taken into account. Within the late--time observational framework adopted in this work, both Gong--Zhang parametrizations remain phenomenologically viable and provide observationally consistent descriptions of the data, with GZ2 exhibiting comparatively tighter constraints and reduced parameter degeneracies.}\\

The marginalized constraints in the $\Omega_m$--$\omega_0$ plane together with the corresponding one--dimensional posterior distributions of $\omega_0$ further clarify the distinct physical behaviour of the two parametrizations. While GZ1 mildly favours quintessence--like behaviour with $\omega_0$ clustered around $-1$ and remaining consistent with $\Lambda$CDM within the $1$--$2\sigma$ region, GZ2 prefers comparatively less negative values of $\omega_0$ together with a noticeably tighter localization of the posterior distribution. This reduced parameter degeneracy is consistent with the comparatively tighter posterior constraints obtained for the GZ2 parametrization and highlights the physical relevance of the Gong--Zhang construction in capturing controlled late--time deviations from $\Lambda$CDM. A complementary $\Omega_m$--$\sigma_8$ analysis further shows that both Gong--Zhang parametrizations remain phenomenologically consistent with the observed late--time structure-growth history after inclusion of the $f\sigma_8$ likelihood, with the GZ2 model again exhibiting comparatively tighter and more localized clustering-sector constraints, indicating a reduced degeneracy between the matter density and clustering amplitude relative to GZ1.\\

Beyond the background expansion history, we have investigated the kinematic and dynamical implications of the Gong--Zhang models through a detailed cosmographic analysis together with the reconstructed evolution of the dark--energy equation of state and the coincidence parameter. Both parametrizations predict a stable late--time accelerated expansion characterized by negative deceleration and well--constrained higher--order cosmographic parameters, confirming that the inferred expansion history remains smooth and free from pathological behaviour. The reconstructed equation--of--state evolution shows that the GZ1 parametrization remains close to a quintessence--like behaviour with $\omega(z)\simeq-1$ near the present epoch and gradually approaches $\omega(z)\rightarrow0$ at high redshift, while GZ2 exhibits a comparatively stronger but still smooth dynamical evolution away from the cosmological--constant limit. Consistent with this behaviour, the coincidence parameter increases progressively toward high redshift in both parametrizations, demonstrating an approximate recovery of an effective matter--dominated expansion behaviour with a dynamically suppressed dark--energy contribution at early times. While the GZ1 evolution remains particularly close to $\Lambda$CDM, the GZ2 parametrization exhibits comparatively stronger but still controlled late--time deviations. The corresponding sound--speed analysis further identifies broad regions in parameter space satisfying the physical stability condition $0<c_s^2<1$, thereby ensuring perturbative stability and causal propagation of dark--energy fluctuations across the observable redshift range. Since the sound speed governs the response of dark energy to inhomogeneities and regulates the propagation of pressure perturbations, these stability regions provide an important microphysical consistency check on the viability of the Gong--Zhang framework.\\

Motivated by this, we next explored the imprint of dynamical dark energy on the growth of cosmic structures through the relative growth--rate deviation with respect to $\Lambda$CDM. Both Gong--Zhang parametrizations accurately reproduce the standard early--time growth behaviour during the matter--dominated era, while exhibiting a mild suppression of structure growth at late times due to the influence of the evolving dark--energy equation of state. The corresponding growth--rate deviations remain smooth and free from instabilities, indicating that the Gong--Zhang models modify late--time gravitational clustering in a controlled and physically consistent manner. This behaviour is fully compatible with current large--scale structure observations and establishes a direct connection between the perturbative stability inferred from $c_s^2$ and the observable evolution of matter fluctuations.\\

Finally, we investigated the thermodynamic imprint of dynamical dark energy through the evolution of the configuration--entropy production rate. The entropy--production--rate deviation $\Delta R(a)$ remains nearly indistinguishable from $\Lambda$CDM during the early matter--dominated epoch, while gradually developing negative deviations at late times as accelerated expansion suppresses the efficiency of gravitational clustering. Both GZ1 and GZ2 exhibit smooth entropy evolution characterized by a late--time minimum followed by a mild recovery near the present epoch, reflecting the competition between dark--energy domination and residual matter clustering. The synchronized behaviour of $\Delta R(a)$ across different conserved mass scales further demonstrates that the entropy evolution is governed primarily by the modified cosmic expansion history rather than by the specific choice of mass scale. These results establish configuration entropy as a complementary thermodynamic probe capable of capturing integrated aspects of cosmic structure formation beyond conventional background and growth observables.\\

{\color{darkblue} Taken together, our results demonstrate that minimal one--dimensional dark--energy parametrizations with physically motivated high--redshift behaviour, such as the Gong--Zhang models, are capable of providing observationally consistent descriptions of current late--time cosmological datasets while simultaneously offering a coherent picture of background expansion, structure growth, perturbative stability, and entropy evolution. Among the two parametrizations, the Type~II model generally exhibits comparatively tighter parameter constraints, reduced degeneracies, and more localized posterior distributions within the observational framework considered in this work. It should be emphasized, however, that the present analysis is based exclusively on late--time cosmological probes and does not incorporate the full Planck CMB likelihood. Consequently, the conclusions of this work should be interpreted within the context of late--time observations only, and future analyses including CMB information may further refine or modify the inferred constraints and model comparison results. Looking ahead, the inclusion of future weak--lensing observations, full DESI measurements, Rubin--LSST supernova samples, and possible scalar--field or modified--gravity realizations of the Gong--Zhang forms may further clarify the physical origin and cosmological implications of these parametrizations. More broadly, the configuration--entropy framework developed in this work opens an alternative thermodynamic perspective on cosmic acceleration and provides a physically motivated criterion for assessing the viability of dynamical dark--energy models beyond conventional background cosmology.}\\

\appendix

\section{CPL parametrization: observational constraints}
\label{app:CPL}

For comparison with the Gong--Zhang dark--energy parametrizations, we additionally consider the Chevallier--Polarski--Linder (CPL) parametrization \cite{Chevallier:2000qy}, characterized by the equation of state $
\omega(z)=\omega_0+\omega_a\frac{z}{1+z}$,
where $\omega_0$ denotes the present value of the dark--energy equation of state and $\omega_a$ quantifies its dynamical evolution with redshift. Assuming a spatially flat FLRW Universe, the corresponding normalized Hubble expansion rate is given by
\[
E^2(z)=\Omega_m(1+z)^3+\Omega_r(1+z)^4
+\left(1-\Omega_m-\Omega_r\right)
(1+z)^{3(1+\omega_0+\omega_a)}
\exp\!\left(-\frac{3\omega_a z}{1+z}\right).
\]

\begin{figure*}[t]
    \centering
    \includegraphics[width=\textwidth]{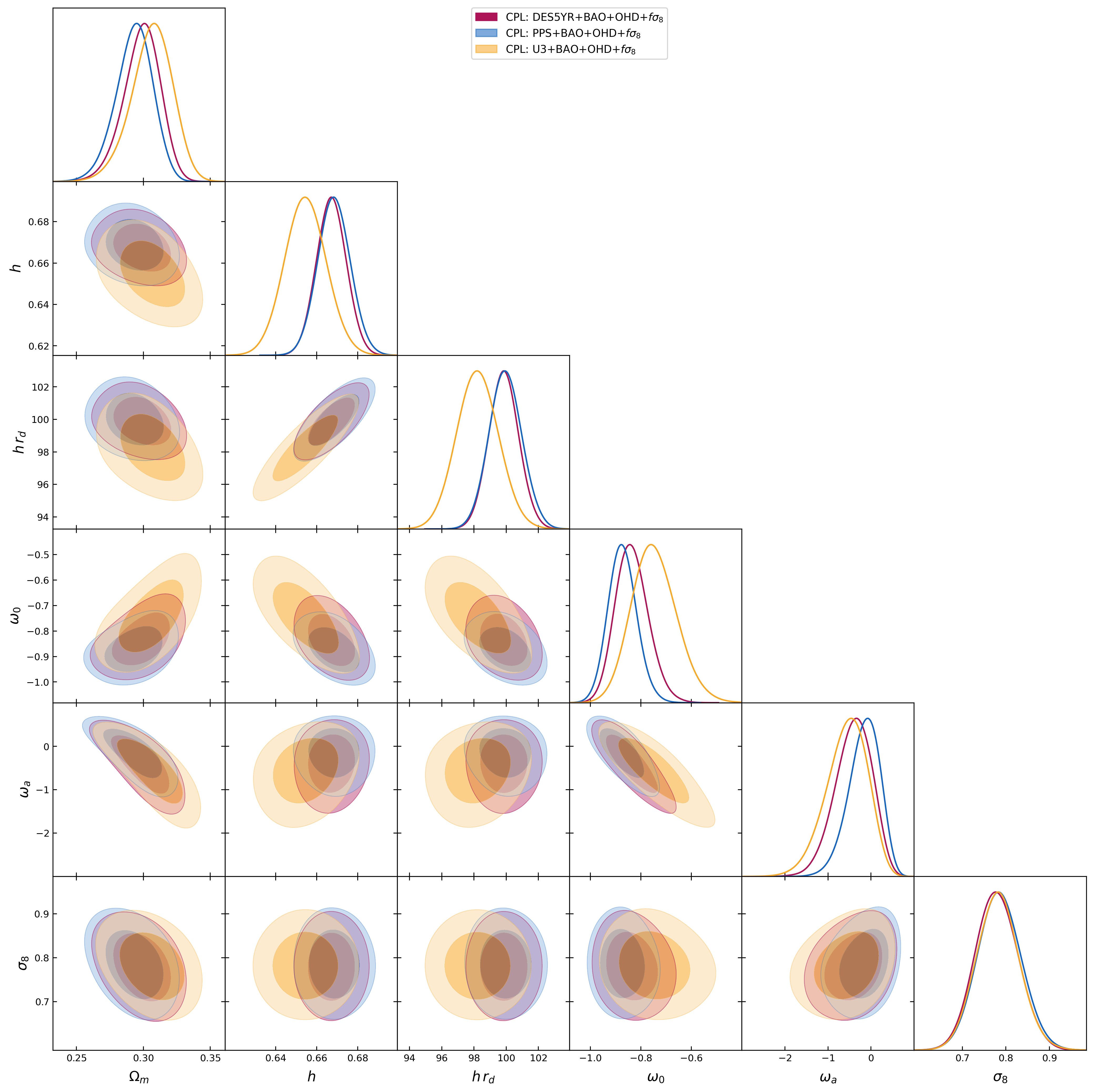}
    \caption{Marginalized one-- and two--dimensional posterior distributions obtained from the Union3+BAO+OHD+$f\sigma_8$, Pantheon+SH0ES+BAO+OHD+$f\sigma_8$, and DES--SN5YR+BAO+OHD+$f\sigma_8$ dataset combinations for the CPL parametrization.}
    \label{fig:CPL_corner}
\end{figure*}

The CPL analysis is included primarily as a benchmark comparison with the restricted one-dimensional GZ1 parametrization. Figure~\ref{fig:CPL_corner} presents the marginalized one-- and two--dimensional posterior distributions for the CPL parameter set
$\{\Omega_m,\,h,\,h\,r_d,\,\omega_0,\,\omega_a,\,\sigma_8\}$
obtained from the Union3+BAO+OHD+$f\sigma_8$, PPS+BAO+OHD+$f\sigma_8$, and DES--SN5YR+BAO+OHD+$f\sigma_8$ dataset combinations. The posterior distributions exhibit the expected correlations among the matter density, Hubble parameter, and dark--energy equation--of--state parameters. In particular, a pronounced anti--correlation between $\omega_0$ and $\omega_a$ is observed, reflecting the intrinsic degeneracy associated with the CPL evolution. Compared to the Gong--Zhang parametrizations, the CPL model exhibits broader posterior regions and comparatively stronger parameter degeneracies, indicating a weaker localization of the dark--energy dynamics within the available late--time observational constraints.\\

\paragraph*{Acknowledgement} This work is supported in part by the National Key Research and Development Program of China under Grant No. 2020YFC2201504. Authors are thankful to the referees for the valuable  suggestions.\\

\paragraph*{Data Availability Statement} All original datasets used in this work are publicly accessible and have been appropriately cited in the bibliography.

\end{document}